\RequirePackage{fix-cm}

\documentclass[pdftex,twocolumn,epjc3]{svjour3}          

\RequirePackage[T1]{fontenc}

\smartqed  

\RequirePackage{graphicx}
\RequirePackage{flushend}
\RequirePackage[numbers,sort&compress]{natbib}
\RequirePackage[colorlinks,citecolor=blue,urlcolor=blue,linkcolor=blue]{hyperref}
\RequirePackage{subfigure}

\journalname{-}

\usepackage{rotating}
\usepackage{amsmath}
\usepackage{amssymb}
\usepackage{mathtools}
\usepackage{multirow}
\usepackage{dblfloatfix}
\usepackage{doi}
\usepackage{xspace}
\usepackage{xcolor}
\usepackage{color}
\usepackage{float}
\usepackage{microtype}
\usepackage[switch]{lineno} 
\usepackage[shortlabels]{enumitem}
\usepackage[toc,page]{appendix}
\usepackage{tikz}
\usetikzlibrary{fit}
\usetikzlibrary{positioning}
\usetikzlibrary{shapes,arrows} 
\usetikzlibrary{calc} 

\definecolor{greyUD}{RGB}{22, 44, 80}
\definecolor{lightBlueUD}{RGB}{0, 70, 255}
\definecolor{greenUD}{RGB}{0, 110, 0}
\definecolor{redUD}{RGB}{255, 0, 0}
\let\oldequation\equation
\let\oldendequation\endequation

\renewenvironment{equation}
  {\linenomathNonumbers\oldequation}
  {\oldendequation\endlinenomath}
\def\antinu{$\bar{\nu}_e$\xspace}
\def\udtc{$^{235}$U\xspace}
\def\udto{$^{238}$U\xspace}
\def\pudtn{$^{239}$Pu\xspace}
\def\pudqu{$^{241}$Pu\xspace}
\def\bm{$\beta^-$\xspace}
\def\ed{$\times 10^{2}\xspace$}
\def\et{$\times 10^{3}\xspace$}
\def\eq{$\times 10^{4}\xspace$}
\def\ec{$\times 10^{5}\xspace$}

\begin{document}
\renewcommand{\arraystretch}{1.3}
\title{Analysis of reactor burnup simulation uncertainties for antineutrino spectrum prediction}

\author{\centering A. Barresi\thanksref{e}
        \and
        M. Borghesi\thanksref{e}
        \and
        A. Cammi\thanksref{j,e1}
        \and 
        D. Chiesa\thanksref{e,e3}
        \and
        L. Loi\thanksref{j,e2}
        \and
        M. Nastasi\thanksref{e}
        \and
        E. Previtali\thanksref{e}
        \and
        M. Sisti\thanksref{e}
        \and
        S. Aiello\thanksref{a}
        \and
        G. Andronico\thanksref{a}
        \and
        V. Antonelli\thanksref{b}
        \and
        D. Basilico\thanksref{b}
        \and
        M. Beretta\thanksref{b}
        \and
        A. Bergnoli\thanksref{c}
        \and
        A. Brigatti\thanksref{b}
        \and
        R. Brugnera\thanksref{c}
        \and
        R. Bruno\thanksref{a}
        \and
        A. Budano\thanksref{d}
        \and
        B. Caccianiga\thanksref{b}
        \and
        V. Cerrone\thanksref{c}
        \and
        R. Caruso\thanksref{a}
        \and
        C. Clementi\thanksref{f}
        \and
        S. Dusini\thanksref{c}
        \and
        A. Fabbri\thanksref{d}
        \and
        G. Felici\thanksref{g}
        \and
        F. Ferraro\thanksref{b}
        \and
        A. Garfagnini\thanksref{c}
        \and
        M. G. Giammarchi\thanksref{b}
        \and
        N. Giudice\thanksref{a}
        \and
        A. Gavrikov\thanksref{c}
        \and
        M. Grassi\thanksref{c}
        \and
        R. M. Guizzetti\thanksref{c}
        \and
        N. Guardone\thanksref{a}
        \and
        B. Jelmini\thanksref{c}
        \and
        C. Landini\thanksref{b}
        \and
        I. Lippi\thanksref{c}
        \and
        S. Loffredo\thanksref{d}
        \and
        P. Lombardi\thanksref{b}
        \and
        C. Lombardo\thanksref{a}
        \and
        F. Mantovani\thanksref{h,i}
        \and
        S. M. Mari\thanksref{d}
        \and
        A. Martini\thanksref{g}
        \and
        L. Miramonti\thanksref{b}
        \and
        M. Montuschi\thanksref{h,i}
        \and
        D. Orestano\thanksref{d}
        \and
        F. Ortica\thanksref{f}
        \and
        A. Paoloni\thanksref{g}
        \and
        E. Percalli\thanksref{b}
        \and
        F. Petrucci\thanksref{d}
        \and
        G. Ranucci\thanksref{b}
        \and
        A. C. Re\thanksref{b}
        \and
        M. Redchuck\thanksref{c}
        \and
        B. Ricci\thanksref{h,i}
        \and
        A. Romani\thanksref{f}
        \and
        P. Saggese\thanksref{b}
        \and
        G. Sava\thanksref{a}
        \and
        A. Serafini\thanksref{c}
        \and
        C. Sirignano\thanksref{c}
        \and
        L. Stanco\thanksref{c}
        \and
        E. Stanescu Farilla\thanksref{d}
        \and
        V. Strati\thanksref{h,i}
        \and
        M. D. C. Torri\thanksref{b}
        \and
        A. Triossi\thanksref{c}
        \and
        C. Tuvè\thanksref{a}
        \and
        C. Venettacci\thanksref{d}
        \and
        G. Verde\thanksref{a}
        \and
        L. Votano\thanksref{g}}

\thankstext{e1}{\textbf{Corresponding author} antonio.cammi@polimi.it}
\thankstext{e2}{lorenzo.loi@polimi.it}
\thankstext{e3}{davide.chiesa@mib.infn.it}

\institute{INFN, Sezione di Milano Bicocca e Dipartimento di Fisica Università di Milano - Bicocca, Italy \label{e} 
\and
INFN, Sezione di Milano Bicocca e Dipartimento di Energia, Politecnico di Milano, Italy \label{j}
\and 
INFN, Sezione di Catania e Università di Catania, Dipartimento di Fisica e Astronomia, Italy \label{a}
\and 
INFN, Sezione di Milano e Università degli studi di Milano, Dipartimento di Fisica, Italy \label{b} 
\and
INFN, Sezione di Padova e Università di Padova, Dipartimento di Fisica e Astronomia, Italy \label{c} 
\and 
INFN, Sezione di Roma 3 e Università degli Studi di Roma Tre, Dipartimento di Fisica e Matematica, Italy \label{d} 
\and 
INFN, Sezione di Perugia e Università degli Studi di Perugia, Dipartimento di Chimica, Biologia e Biotecnologie, Italy \label{f} 
\and
Laboratori Nazionali dell'INFN di Frascati, Italy \label{g}
\and
INFN, Sezione di Ferrara, Italy \label{h}
\and
Università degli Studi di Ferrara, Dipartimento di Fisica e Scienze della Terra, Italy \label{i}}

\date{Received: 05 March 2024 - Accepted: 25 September 2024}
\twocolumn
\maketitle
\begin{abstract}{
Nuclear reactors are a source of electron antineutrinos due to the presence of unstable fission products that undergo $\beta^-$ decay.
They will be exploited by the JUNO experiment to determine the neutrino mass ordering and to get very precise measurements of the neutrino oscillation parameters. This requires the reactor antineutrino spectrum to be characterized as precisely as possible  both through high resolution measurements, as foreseen by the TAO experiment, and detailed simulation models.
In this paper we present a benchmark analysis utilizing Serpent Monte Carlo simulations in comparison with real pressurized water reactor spent fuel data. Our objective is to study the accuracy of fission fraction predictions as a function of different reactor simulation approximations.
Then, using the BetaShape software, we construct reactor antineutrino spectrum using the summation method, thereby assessing the influence of simulation uncertainties on it.}

\end{abstract}

\section{Introduction}
\label{sec:introduction}
The knowledge of the electron antineutrino (\antinu) flux emitted by nuclear reactors is of crucial importance for the Jiangmen Underground Neutrino Observatory (JUNO)~\cite{An2016, JUNO_new_ref} experiment, which will exploit 8 reactors as \antinu sources.
The JUNO detector, a 20-thousand-ton liquid scintillator located at $\sim$ 53~km from the Yangjiang and Taishan Nuclear Power Plants, is in its final construction phase in China.

The reactor \antinu will be detected through the Inverse Beta Decay (IBD) reaction: 
\begin{equation}
    \bar{\nu}_e + p \xrightarrow[]{} n + e^+
\end{equation}
which has a \antinu energy threshold of 1.8~MeV.

One of the main goals of the JUNO experiment is to determine the neutrino mass ordering by analyzing the pattern of \antinu oscillations at medium baseline. 
This experimental measurement requires high accuracy, therefore it is mandatory to control all the systematic uncertainties related to the reactor \antinu source.
For this purpose, the Taishan Antineutrino Observatory (TAO)~\cite{TAO} detector will measure with unprecedented energy resolution the non-oscillated \antinu spectrum at $\sim40$~m from one of the reactors exploited by JUNO. The importance of computational tools for predicting reactor antineutrino fluxes stands out in this framework, since in principle, if they are validated with experimental data, they can be used for flux shapes predictions, by knowing only reactor operation information. That could be the case for both the second Taishan reactor core as well as the six reactors of Yangjiang plant, which lack of near field detectors. Also, nuclear database accuracy could be benchmarked against high energy resolution TAO data.

Until now, two different methods have been used to model the reactor \antinu flux: the conversion~\cite{Huber2011,Mueller2011} and the summation 
(also called \textit{ab-initio})~\cite{Fallot2012, Estienne_ab_initio} methods. 
Both approaches are affected by uncertainties related to the input data, leading to a discrepancy in both \antinu flux intensity and spectral shape~\cite{HayesVogel2016, ZHANG2024104106} when compared with the experimental data acquired by near detectors~\cite{DayaBay:2022jnn,DoubleChooz:2019qbj,RENO:2015ksa, Stereo2023, Ko_2017, Andriamirado_2023}.

Since JUNO and TAO will feature an unprecedented energy resolution in the field of \antinu detectors, the summation method is the most suitable to use, because it gives access to the fine structure of the spectrum~\cite{Dwyer2015}. 

To determine the real time reactor \antinu flux and spectrum emitted by a nuclear reactor two ingredients, completely independent of each other, are needed:

\begin{enumerate}
    \item the activity of \bm decaying nuclides in the reactor core as a function of time;
    \item the \antinu spectrum characteristics of each \bm decaying nuclide.
\end{enumerate}

In this paper we present the development of an analysis tool designed explicitly for probing the uncertainties inherent in the first ingredient. We choose as test model a standard PWR system due to the data availability of most of its components. Even if the reactor design is not strictly equal to the reactors placed in Taishan and Yangjiang, the presented results are representative of a light water reactor burnup cycle. In particular, the main differences in the present work are a different \udtc assembly enrichment and a lower nominal power with respect to the two Chinese plants. This work is meant to show the methodological approach to reproduce the antineutrino spectrum from a reactor model, highlighting some features that can have an impact on it. Future works will be devoted to have a high fidelity simulation of both the plants involved in JUNO-TAO. Through meticulous simulation of reactor neutronics and fuel burnup, we attain a comprehensive understanding of the production and decay processes of all contributing nuclides that shape the \antinu flux. In a second step we couple reactor burnup simulations to an extensive  \antinu spectral database to investigate the reactor-related uncertainties affecting the non-oscillated reactor \antinu spectra.
This aspect can be considered of extreme importance for all reactor neutrino experiments since it gives the key to reactor \antinu flux control, allowing its real time spectral reconstruction and providing a continuous monitoring of the \antinu flux variations. 

This paper is organised as follows: 
\begin{itemize}
    \item in Sect.~\ref{sec:methodology} we describe the methodology and the tools that we use to perform our analysis;
    \item in Sect.~\ref{sec:validation} we perform a benchmark analysis with the experimental data of a power reactor, to validate the Monte Carlo simulations of fuel burnup at different levels of approximation;
    \item in Sect.~\ref{sec:antinu_spectra} we compare the fission fractions and the non oscillated reactor \antinu spectrum at different burnup levels obtained with different approximations in the reactor simulations;
    \item in Sect.~\ref{sec:conclusions}, summary and future works are pointed out.
\end{itemize}

\section{Methodology and tools}
\label{sec:methodology}
During the operational phase of a Pressurized Water Reactor (PWR), thermal power $W_{th}$ is continuously produced through fission reactions. In the fresh fuel, usually made of uranium dioxide (UO$_2$) enriched in \udtc between 2 wt\% and 5 wt\%, only uranium isotopes (\udtc and \udto) can be fissioned by neutrons. However, as reactor operation proceeds in time, a series of neutron capture reactions and radioactive decays leads to accumulation of \pudtn and \pudqu, which are fissile by thermal neutrons.
The fission fragments are nuclei characterized by an excess of neutrons, thus they undergo a series of \bm decays to reach nuclear stability. On average, about 6 \antinu are emitted by \bm decays after each fission event.
It is therefore possible to consider each fissile element in the reactor as the parent of several unstable fission products. 

\begin{figure*}[htbp]
\centering
\includegraphics[width=1\linewidth]{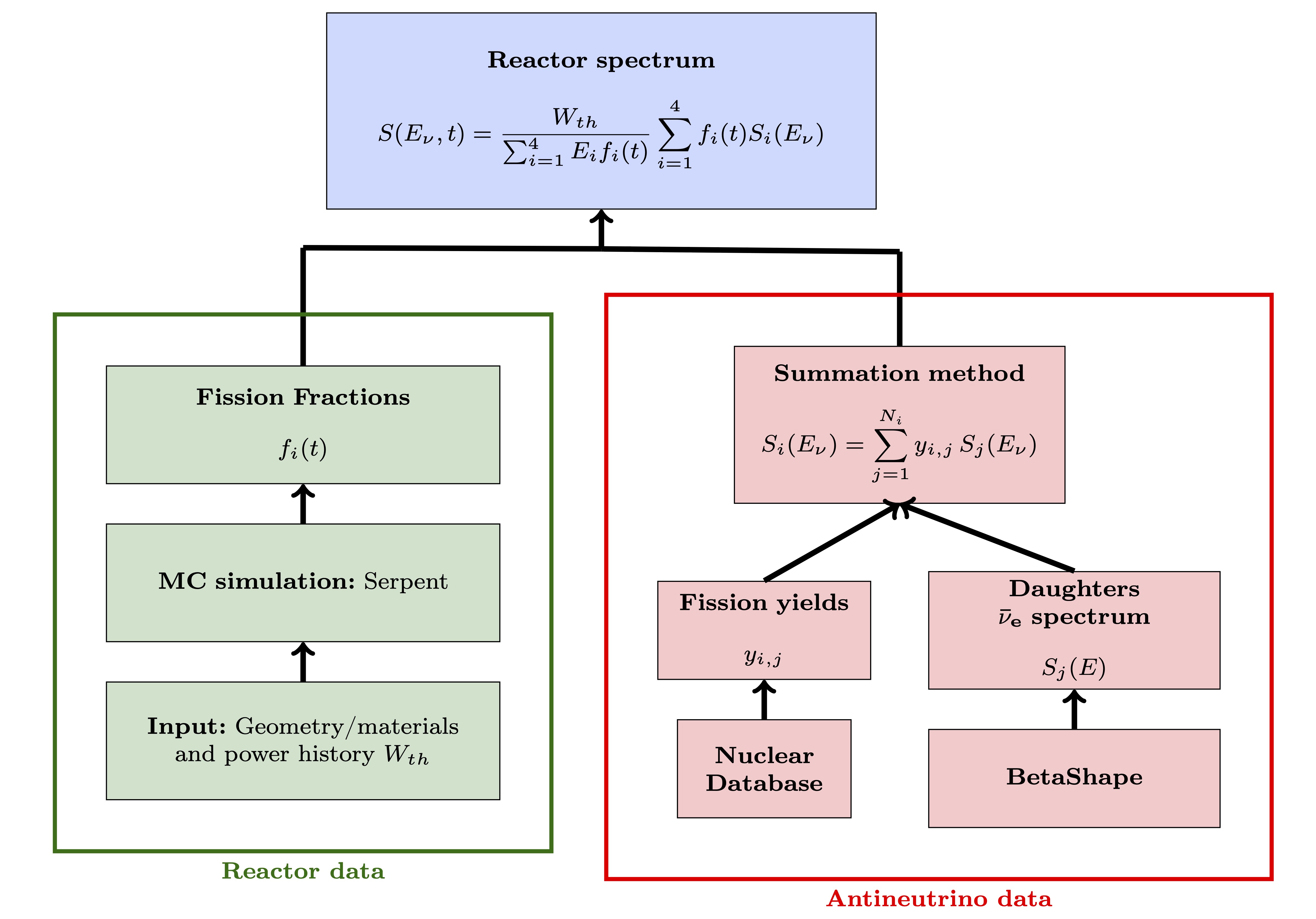}
\caption{Scheme for the calculation of the reactor antineutrino spectrum. Green box: starting from accessible plant information, the Monte Carlo tool predicts the evolution of fission fractions. The importance of this methodology lies in the flexibility with which input parameters can be varied to accommodate daily power profiles or different reactor designs. Red box: we exploit spectra computed with the BetaShape tool along with the cumulative fission yields to evaluate the fissile spectra at equilibrium with the summation method.}
\label{fig:MethodologyScheme}
\end{figure*}

The resulting spectrum of \antinu 
emitted per unit time by a nuclear reactor $ S(E_{\nu},t)$ can be quantified through the following formula: 
\begin{equation}
\label{eqn:SpectrumEquationI}
    S(E_{\nu},t) =\frac{W_\text{th}(t)}{\sum_{i=1}^4 E_i f_i (t)} \sum_{i=1}^{4} f_i (t) S_i(E_{\nu},t),
\end{equation}
where $i$ labels the main fissile nuclides in PWR reactors (\udtc, \udto, \pudtn, \pudqu), $W_{th}(t)$ is the reactor thermal power, $E_i$ is the average energy deposited in the reactor per fission of the $i$-th nuclide, $f_i(t)$ are the \textit{fission fractions} (i.e. the ratio of the fission rate of the $i$-th nuclide to the total fission rate). Finally, $S_i(E_{\nu},t)$ is the spectrum associated to the fission of the $i$-th nuclide estimated with the \textit{summation method}, where the antineutrino spectra of the daughters of $i$ are summed together, each with a weight that reflects their decay rate in the system. It is worth to highlight that the time dependence in $S_i(E_{\nu},t)$ will not be considered in this work because of the equilibrium hypothesis (see Sec. \ref{sec:AntineutrinoParameters}). This choice is made to underline the effects of reactor parameters on the fission fractions, according to the most used approach in the international community adopting ab-initio scheme. Non-equilibrium studies will be faced in future works.\\
Equation \eqref{eqn:SpectrumEquationI} highlights that the reactor \antinu spectrum depends on two sets of parameters: the first one, including $f_i(t)$ and $W_\text{th}(t)$, is related to reactor history, power and fuel burnup. The second comprises the nuclear data used to extract the different \antinu spectra from the corresponding \bm decays. 
In the following subsections, we describe the tools that we use to: \begin{itemize}
    \item calculate reactor-related parameters (\ref{sec:ReactorParameters}) using the flexible Serpent (version 2.1.31) environment for fissile system simulations;
    \item build the \antinu spectra of the four actinides (\ref{sec:AntineutrinoParameters}) adopting the summation method, with the recently developed BetaShape (version 2.3) code. 
\end{itemize}
These tools will be subsequently coupled according to the scheme reported in Fig.~\ref{fig:MethodologyScheme}, to get the reactor \antinu spectrum.

\subsection{The Serpent tool for reactor simulations}
\label{sec:ReactorParameters}

Neutron-matter interactions occurring within a fissile system (i.e. a nuclear reactor) can be described by the coupling of two equations:
\begin{itemize}
    \item Steady State Boltzmann equation~\cite{BellGlasstone}: by knowing the concentration of all nuclides in the system (nuclide field) $N(\mathbf{r},t)$, this equation solves the criticality problem in the state space $\mathbf{s} = (\mathbf{r} , \mathbf{\Omega}, E, t)$ for the angular neutron flux $\varphi$ with a balance between the neutrons produced by fission and the neutrons which disappear by leakage or absorption:
    \begin{equation}
    \label{eqn:BoltzmannEquation}
         \text{L}(N)\varphi(\mathbf{s}) - \frac{1}{k_{\text{eff}}}\text{F}(N) \varphi(\mathbf{s}) = 0,
    \end{equation}
    where $\text{L}(N)\varphi(\mathbf{s})$ represents the migration and loss of neutrons from $\mathbf{s}$, $\text{F}(N)\varphi(\mathbf{s})$ accounts for the neutron production in $\mathbf{s}$ due to fission and $k_{\text{eff}}$ is the multiplication factor. 
    \item Bateman equation~\cite{Bateman}: it describes the time evolution of a nuclide field subjected to a neutron flux. This equation (also called \textit{burnup equation}) takes into account both the fuel depletion and the transmutation processes in the system. The Bateman equation reads as an Ordinary Differential Equation (ODE) system:
    \begin{equation}
    \label{eqn:BatemanEquation}
    \frac{\text{d}N(\mathbf{r},t)}{\text{d}t} = \left[ \int_0^{\infty} \phi(\mathbf{r}, E, t)\mathbb{X}(E,T) \, \text{d}E + \mathbb{D} \right]N(\mathbf{r},t) ,
    \end{equation}
    where $N(\mathbf{r},t)$ is the array of nuclide concentrations, $\phi(\mathbf{r}, E ,t) = \int_{4\pi} \varphi(\mathbf{s}) \, \text{d}\mathbf{\Omega}$ is the scalar neutron flux at time $t$, $\mathbb{X}(E,T)$ is the cross-section (dependent on the neutron energy $E$ and the local material temperature $T$) and fission yield matrix, and $\mathbb{D}$ is the decay matrix, which accounts for the free radioactive evolution. 
\end{itemize}
Many codes are able to solve the neutron transport problem described by Eq.~\eqref{eqn:BoltzmannEquation} using the Monte Carlo (MC) method.
Some MC codes, specialized for reactor simulations, are supplemented by burnup sequence schemes that allow to solve the coupled system described by Eq.~\eqref{eqn:BoltzmannEquation} and \eqref{eqn:BatemanEquation}. 
The latter equation has a formal solution only when the neutron flux is constant in time. 
However, in reality, the neutron flux is a time-dependent quantity because it varies depending on the nuclide field distribution. 
In order for the analytical solution to be used, the time frame is divided in intervals in which the flux is treated as a time-constant variable. Several methods are given in the literature for choosing the constant value to  be used within
each interval~\cite{BurnupMethods}.

The three-dimensional continuous-energy Monte Carlo code Serpent~\cite{SERPENT} was chosen for this work. This analysis tool, created by the Finnish research Centre VTT, is capable of simulating the neutronics and fuel evolution of nuclear reactors with a great level of detail, therefore allowing to estimate the neutron flux and,  subsequently, derive the reactions rates. In particular, it is possible to compute:
\begin{equation}
\label{eqn:ReactionRate}
    R = \frac{1}{V} \int_V \int_{4\pi}\int_{E_{g-1}}^{E_g} h(\mathbf{r}, E) \phi (\mathbf{r}, E) \, \text{d}\mathbf{r}\,\text{d}\mathbf{\Omega}\,\text{d}E,
\end{equation}
where $V$ and $E_{g-1},E_{g}$ are, respectively, the spatial and energy domains in which the neutron flux is estimated, while $h(\mathbf{r}, E)$ is a generic \textit{response function}. The choice of $h$ allows different quantities to be estimated during the neutron transport simulation. 
In the framework of this work, the most used response functions are:
\begin{itemize}
    \item $h = 1$ to estimate the volume averaged neutron flux;
    \item $h = n_i(\mathbf{r})\sigma_{f,i}(E,T)$, where $n_i(\mathbf{r})$ is the atomic density of the $i$-th nuclide and $\sigma_{f,i}(E,T)$ its microscopic fission cross section, to estimate the fission reaction rates per unit volume;
    \item $h = \sum_{k=1}^4 n_k(\mathbf{r})\sigma_{f,k}(E,T)$ to get the total fission reaction rate per unit volume.
\end{itemize}
Within the reactor \antinu analysis framework, the integration ranges to be used in Eq.~\eqref{eqn:ReactionRate} to get the fission fractions $f_i$ are the whole energy spectrum and the entire spatial domain: 
\begin{equation}
\label{eq:FF}
     f_i = \frac{\int_V \int_0^\infty n_i(\mathbf{r})\sigma_{f,i}(E,T) \phi (\mathbf{r}, E) \, \text{d}\mathbf{r} \, \text{d}E }{\sum_{k=1}^4 \int_V \int_0^\infty n_k(\mathbf{r})\sigma_{f,k}(E,T) \phi (\mathbf{r}, E) \,\text{d}\mathbf{r} \, \text{d}E}      ,
\end{equation}

\subsection{Antineutrino spectra production}
\label{sec:AntineutrinoParameters}
As previously pointed out, each fissile nuclide $i$ can be associated with a corresponding \antinu spectrum $S_i(E_{\nu},t)$ given by the decays of its fission products. 
The spectrum, normalized, can be written as:
\begin{equation}
\label{eqn:ActinideSpectra}
    S_i(E_{\nu}, t) = \frac{\sum_{j=1}^{N_i} A_{i,j}(t) \, S_j(E_{\nu})}{ \sum_{j=1}^{N_i} A_{i,j}(t) \int_0^\infty \, S_j(E_{\nu}) \text{d}E_{\nu} },
\end{equation}
where $A_{i,j}(t)$ is the activity of the nuclide $j$ resulting from the fission of the nuclide $i$, and $S_j(E_{\nu})$ is the \antinu spectrum associated to the decay of the fission product $j$. 

Even if it is possible to get $A_{i,j}(t)$ from Serpent simulations, in this work we decided to build the \antinu spectra at equilibrium (i.e. under the condition that the production rate of a nuclide equals its disappearance), thus removing the time dependence in Eq.~\eqref{eqn:ActinideSpectra}. 
Since transmutation processes due to neutron capture are relevant only for radionuclides with $\beta^-$ Q-value below the IBD energy threshold (1.8~MeV), we can neglect their contribution assuming that, at equilibrium:
\begin{equation}
    A_{i,j} \propto y_{i,j},
\end{equation}
where $y_{i,j}$ is the cumulative fission yield (CFY), i.e. the probability of production of nuclide $j$ per each fission of $i$. 
Moreover, it is worth noting that most part of \bm decays with Q-values higher than 1.8~MeV are characterized by half-lives much shorter than the typical reactor steady-state operation time, thus the equilibrium condition is quickly reached for most nuclides emitting \antinu detectable through the IBD reaction.
The adoption of the equilibrium approximation greatly simplifies the computation of $S_i(E_{\nu})$, since the values of $y_{i,j}$ and their uncertainties are tabulated in nuclear databases for fission induced by thermal, and fast neutrons. The correlations affecting the fission yield uncertainties are not included in this analysis since a complete database is currently missing.

The other ingredients needed to build the \antinu spectrum associated to the $i$-th nuclide are the \antinu spectra emitted by its fission products. For this purpose, we decided to exploit the BetaShape code, developed at the Laboratoire National Henri Bequerel (LNHB). This code has been developed to improve the accuracy in the analytical calculations of beta spectra. 
BetaShape takes as input the Evaluated Nuclear Structure Data Files (ENSDF) published in various nuclear databases and provides as outputs the electron and \antinu spectra of any \bm transitions.
More details on the assumptions and approximations adopted in BetaShape can be found in~\cite{BetaShape_ANewCode}. 

\section{Validation of reactor simulations}
\label{sec:validation}
In order to validate the Monte Carlo simulations that we use to compute the fission fractions, we performed a benchmark analysis of simulation results with available experimental measurements.
Since direct fission fractions measurements are not possible from an operational point of view, we adopt an indirect approach, validating our burnup simulations with published measurements of nuclide concentrations in the spent fuel of a PWR reactor.
Indeed, as shown in Eq.~\ref{eq:FF}, the fission fraction $f_i(t)$ depends on the concentrations $n_k(\mathbf{r})$ of the four fissile nuclides.
If we neglect the uncertainties related to the microscopic cross section (which will be addressed in future analyses), it is possible to conclude that the accuracy in the estimation of the fission fractions can be traced back to the ability to reproduce the nuclide field (or the neutron flux, since these two parameters are connected by the burnup scheme).
 
During the past decades, experimental campaigns have been carried out in order to obtain nuclide concentration data of irradiated fuels, through destructive radiochemical techniques~\cite{PIE_ORNL}. These measurements are considered internationally established and a recognised benchmark.
In this work, we use the Takahama-3 data-set as a test case, comparing the concentrations of the four fissile nuclides sampled in two spent fuel rods with the predictions of our Serpent MC simulations. 
Among the various references in the literature, the Takahama-3 reactor has features very similar, in terms of neutron energy spectrum and fuel type, to those of the reactors that will be used by JUNO and TAO. 
Moreover, its fuel assemblies reached a high level of burnup, thereby allowing for having a larger isotopic inventory, useful for a time-resilient validation of the simulation outcomes.

\begin{figure}[h]
    \centering
    \includegraphics[width=0.48\textwidth]{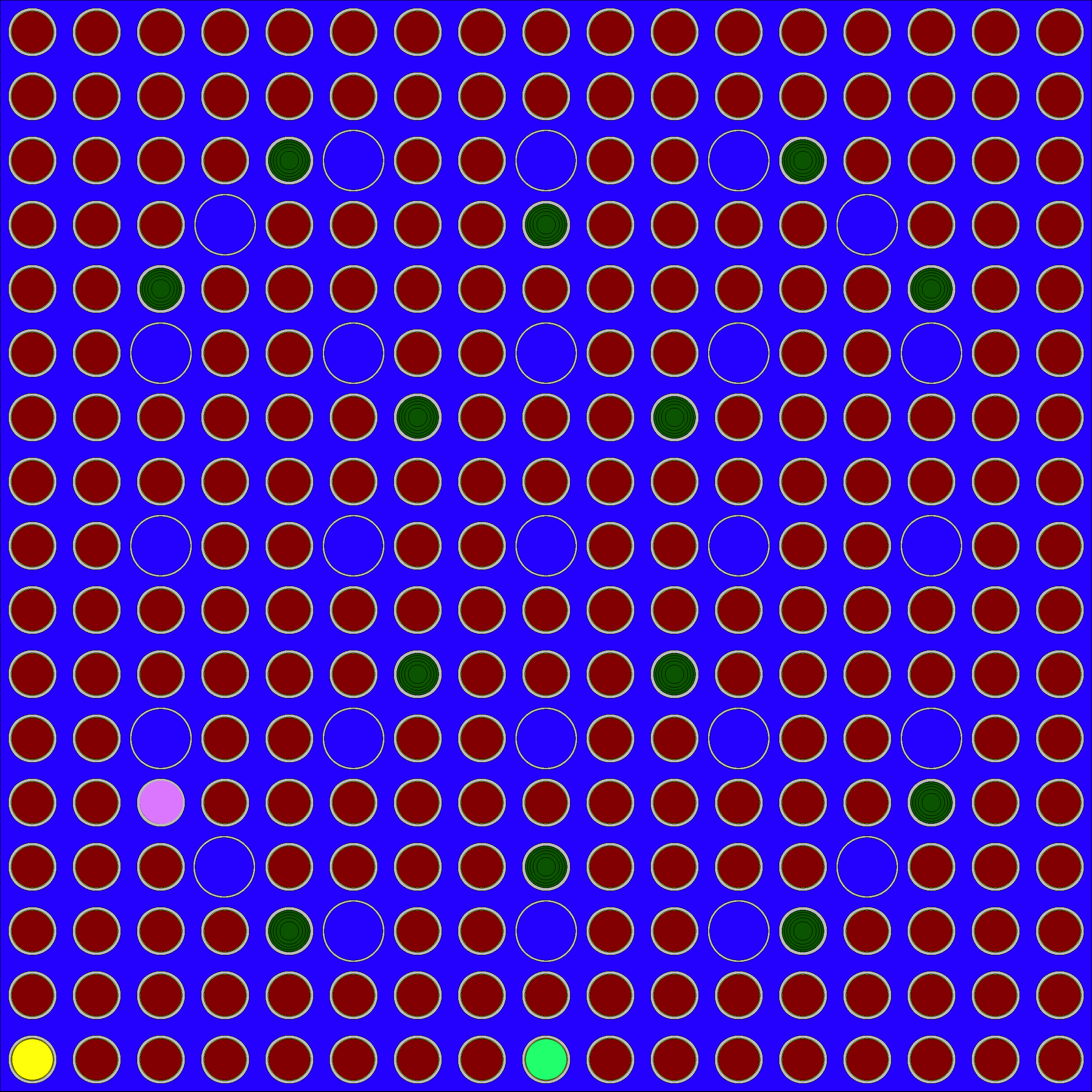}
    \caption[Top view of Takahama Assembly]{Top view of Takahama-3 fuel  assembly design. The yellow circle indicates the position of SF95, the pink one shows the SF96 position, whereas the light green is representative of the SF97 location. The red color is used for the other standard fuel pins, whereas dark green is used for the Gd-loaded ones. Finally, blue represents water. \label{fig:TakahamaAssembly}}
\end{figure} 

The Takahama-3 is a PWR operated by the Kansai Electric Power Company in Japan since 1984~\cite{IAEA_reactor}. 
This reactor is characterized by a nominal thermal power of 2652~MW. 
Its core comprises 157 fuel assemblies, each of which contains a lattice structure of 17$\times$17 rod bundle. Each rod has a radius of $\sim 0.4$~cm, while the side of the squared lattice is 21.4~cm and the assembly total height is about 4~m with $\sim 3.64$~m of active length. 
Each fuel assembly contains 250 pins with standard fuel (i.e. UO$_2$  with an enrichment of $^{235}$U/U = 4.11 wt\%), 14 fuel pins with gadolinium as a burnable absorber ($^{235}$U/U = 2.63 wt\% and a poison ratio of Gd$_2$O$_3$/Fuel = 6.0 wt\%), and 25 guide tubes filled with water in correspondence to the control rod positions.
In particular, spent nuclear fuel analyses were performed on three fuel rods, named SF95, SF96 and SF97, respectively belonging to the NT3G23 assembly (the first two) and to the NT3G24 assembly (the third one).
SF95 and SF97 are standard fuel pins, whereas SF96 is a Gd-loaded fuel pin.
These rods were irradiated in the core for a certain (documented) time; then, after the extraction, five or six samples were drilled out of them at different axial locations and analyzed with Isotope Diluted Mass Spectrometry (IDMS) to measure the concentration of uranium and plutonium isotopes. 
For this analysis, we chose to focus only on the four actinides that build up the fission fractions, since they are the most relevant isotopes in the reactor antineutrino framework. 
Moreover, since the Gd-loaded fuel exhibit a complex evolution due to a non-uniform depletion in the radial direction \cite{radialGd}, and given that the available data refer exclusively to the axial direction, we decided to discuss only the data linked to SF95 and SF97 standard fuel pins, even if we aknowledge that it is an important point to be addressed for EPR and CPR reactors, which include Gd-fuel.

In Fig.~\ref{fig:TakahamaAssembly} we show the position of the rods SF95, SF96 and SF97 in their assemblies and in Table~\ref{tab:GeometricalValuesTakahama} we summarize the available information about the position and the local burnup (estimated through the $^{148}$Nd method~\cite{Nd148_method1, Nd148_method2}) of the samples taken from the rods labelled with a progressive number from highest to lowest position.  

\begin{table*}[ht]
\centering 
    \renewcommand{\arraystretch}{1.2}
    \begin{tabular}{c p{1.5cm} p{1.75cm} p{1.75cm}  p{1.75cm} p{1.75cm} p{1.75cm} p{1.75cm}}
    \hline
      \textbf{Assembly} & \textbf{Sample ID} & \textbf{Axial position (cm)} & \textbf{Burnup (MWd/kg)} & $\bf{^{235}}$\textbf{U (g/tonU)} & $\bf{^{238}}$\textbf{U (g/tonU)} & $\bf{^{239}}$\textbf{Pu (g/tonU)} & $\bf{^{241}}$\textbf{Pu (g/tonU)} \\
      \hline\hline
      \multirow{5}{*}{NT3G23} & SF95-1  &  4.2  &  14.30 & 2.674\eq & 9.499\ec & 4.227\et & 3.69\ed\\
                              & SF95-2  &  20.2  &  24.35 & 1.927\eq & 9.424\ec & 5.655\et & 9.578\ed\\
                              & SF95-3  &  72.2  &  35.52 & 1.326\eq & 9.338\ec & 6.194\et & 1.486\et\\
                              & SF95-4  &  200.2 &  36.69 & 1.23\eq & 9.335\ec & 6.005\et & 1.466\et \\
                              & SF95-5  &  340.2 &  30.40 & 1.544\eq & 9.388\ec & 5.635\et & 1.153\et\\
      \hline
      \multirow{6}{*}{NT3G24} & SF97-1  &  0.40  &  17.69 & 2.347\eq & 9.493\ec & 3.844\et & 4.237\ed \\
                              & SF97-2  &  19.1  &  30.73 & 1.571\eq & 9.377\ec & 5.928\et & 1.235\et \\
                              & SF97-3  &  46.8  &  42.16 & 1.030\eq & 9.282\ec & 6.217\et & 1.689\et \\
                              & SF97-4  &  168   &  47.03 & 8.179\et & 9.246\ec & 6.037\et & 1.77\et \\
                              & SF97-5  &  276.7 &  47.25 & 7.932\et & 9.247\ec & 5.976\et & 1.754\et \\
                              & SF97-6  &  339.7 &  40.79 & 1.016\eq & 9.310\ec & 5.677\et & 1.494\et \\
      \hline
    \end{tabular}
    \caption[Geometric data of Takahama-3 samples]{Data of axial positions and local burnup of the samples drilled from the SF95 and SF97 pins. The axial locations have to be intended as the distance from the top of the fuel, as reported in the original documentation. The local burnup was evaluated with the $^{148}\text{Nd}$ technique, which brings a $\pm3\%$ uncertainty. Finally we report the measured concentrations for \udtc, \udto, \pudtn and \pudqu in units of grams per uranium tons in the fresh fuel. Data from~\cite{Takahama3}.}
    \label{tab:GeometricalValuesTakahama}
\end{table*}

\subsection{Geometry and materials implementation}
\label{sec:GeometryAndMaterials}

In order to simulate the fuel evolution in the SF95 and SF97 pins of the Takahama reactor, we implemented the geometry and the materials of the whole assembly in Serpent, taking as reference the first table in~\cite{Takahama3}. 
Fig.~\ref{fig:TakahamaAssembly} shows the assembly section in the xy plane of the simulated geometry. At the boundaries of the x and y directions, we apply reflective conditions to simulate the presence of the surrounding fuel assemblies. In the z direction, the geometry extends for 4.46~m, including the caps of the fuel cladding and a 20~cm thick water layer at top and bottom, to take into account the reflection of neutrons by water, then, black boundary condition applies. Having the reflective boundary conditions in the x-y dimensions, this makes the assembly to burn as it was placed in the centre of the reactor core. During real cycle operations, assemblies are moved within the core structure when new fuel is added. Since no actual core position information were given, we acknowledge this extra uncertainty during the analysis.
To define the materials in Serpent simulation, we used the cross sections of the ENDF/B-VII.1~\cite{ENDF_B_VII} nuclear library.

\subsection{Power normalization}

\begin{table*}[h]
\centering
    \renewcommand{\arraystretch}{1.5}
    \begin{tabular}{c c c c c c}
    \hline
      \textbf{Start} & \textbf{Stop} & \textbf{Cycle} & \textbf{State} & \textbf{Time interval (d)}& \textbf{Rods} \\
      \hline
      26/01/1990 & 15/02/1991 & 5 & On   & 385 & SF95, SF96, SF97 \\
      15/02/1991 & 14/05/1991 &   & Cool & 88  & SF95, SF96, SF97 \\
      14/05/1991 & 19/06/1992 & 6 & On   & 402 & SF95, SF96, SF97 \\
      19/06/1992 & 20/08/1992 &   & Cool & 62  & SF97 \\
      20/08/1992 & 30/09/1993 & 7 & On   & 406 & SF97 \\
      \hline
    \end{tabular}
    \\[10pt]
    \caption[Irradiation history of Takahama-3 experiment]{Irradiation history of Takahama-3 experiment. Data taken from~\cite{Takahama3}.}
    \label{tab:IrradiationHystoryTakahama}
\end{table*}

\begin{table}[h]
\centering 
    \renewcommand{\arraystretch}{1.5}
    \begin{tabular}{c c c}
    \hline
      & \textbf{SF95}   & \textbf{SF97}  \\
      \hline
     \textbf{Cycle 5} & 0.04576 & 0.037147 \\
     \textbf{Cycle 6} & 0.04013 & 0.039128\\
     \textbf{Cycle 7} & - & 0.034402\\
      \hline
    \end{tabular}
    \\[10pt]
    \caption[Average Specific Power for Takahama-3 experiment]{Average power densities used to normalize the Serpent simulations of the Takahama-3 fuel burnup. Entries are in kW/gU of initially loaded uranium.}
    \label{tab:SpecificPowerTakahama}
\end{table}

In this work we simulate with a good level of approximation the irradiation history of the two Takahama assemblies, considering  both the different power levels during the reactor operation cycles\footnote{Each cycle usually has a duration of $\sim 12-18$ months, after which each assembly can either be removed from the core or moved into another core location.} and the cooling periods.
The technical report cited above (\cite{Takahama3}) provides the power density values at the sampling points as a function of time since reactor start-up (approximately, monthly data). The unit used to express power density is kW/g(U), i.e. kW divided by the mass of Uranium initially loaded in the sampled volume.
From these reactor operation data, it is possible to estimate the average power density for every pin in each depletion cycle, so as to set the correct input power values (used to normalize the neutron fluxes and the reaction rates) in the Serpent simulations.
In particular, the following steps were performed:
\begin{itemize}
    \item an effective local power value was calculated for each sampling point as the average value of the powers in the pin over the time intervals when the reactor was at full power;
    \item after extracting the time-weighted axial power profile for the SF95 and SF97 pins, the integral mean value of the power was calculated to get the effective power density for the normalization.
\end{itemize}
In Table~\ref{tab:IrradiationHystoryTakahama} we report the dates and the time intervals in which the assemblies were in operation or in a cooling phase. Fuel assembly NT3G23 (and thus the fuel pin SF95) was irradiated for two cycles, whereas the assembly NT3G24 (and thus the fuel pin SF97) was irradiated for three cycles, reaching a higher burnup level than the previous one. 
Finally, Table~\ref{tab:SpecificPowerTakahama} lists the average specific power used for the pins in the cycles.

\subsection{Burnup simulation}
\label{sec:BurnupSimulation}

\begin{figure}[h]
    \centering
    \includegraphics[width=0.48\textwidth]{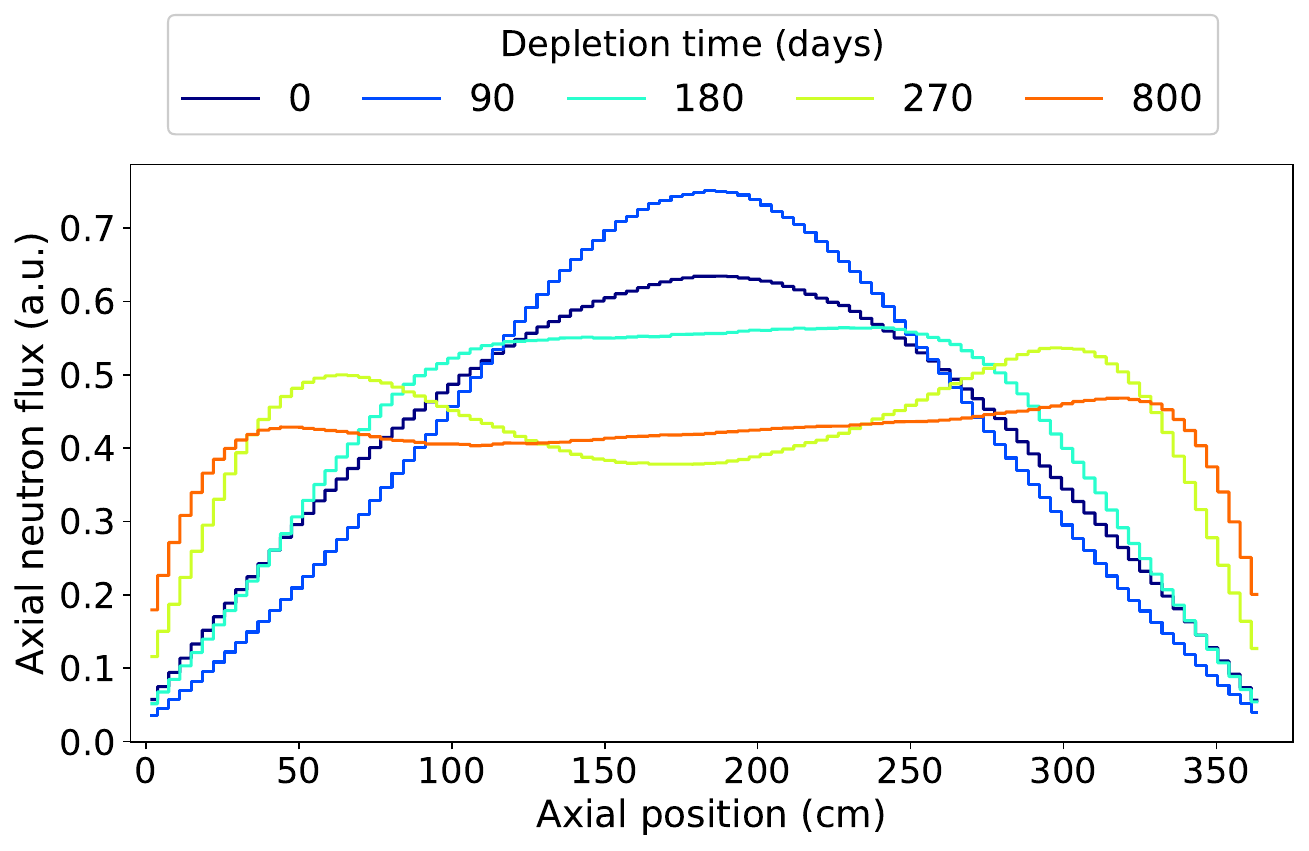}
    \caption[Axial flux]{Axial neutron flux of SF95 rod at several depletion steps, from the start of the cycle up to 800 days. Flux behaviour reflects the fuel evolution, progressively decreasing the number of neutrons in the central regions due to burnup. The zero of the axial position coincide with the top of the rod, accordingly to the Takahama report.}
    \label{fig:FreshFuel_axial}
\end{figure}

The assemblies were simulated during the depletion cycle, starting from the fresh fuel configuration, with the purpose of studying the spatial and energy distributions of neutron fluxes. 
In Fig.~\ref{fig:FreshFuel_axial}, we show the axial profile of the neutron flux intensity at several depletion steps, obtained by integrating on the full energy domain in Eq.~\eqref{eqn:ReactionRate} and a spatial mesh with 100 equal subdivisions from $z =-182.4$ cm to $z=182.4$ cm, corresponding to the volume occupied by the fuel inside the pin. 
When the fuel is fresh, the neutron flux is spatially distributed with a bell shape, having the effect of burning the \udtc atoms faster in the central part of the rod. As a consequence, as the fuel depletion proceeds, the axial profile becomes more flat, progressively equalizing the nuclide concentrations along the z axis.

\begin{figure}[h]
    \centering
    \includegraphics[width=0.48\textwidth]{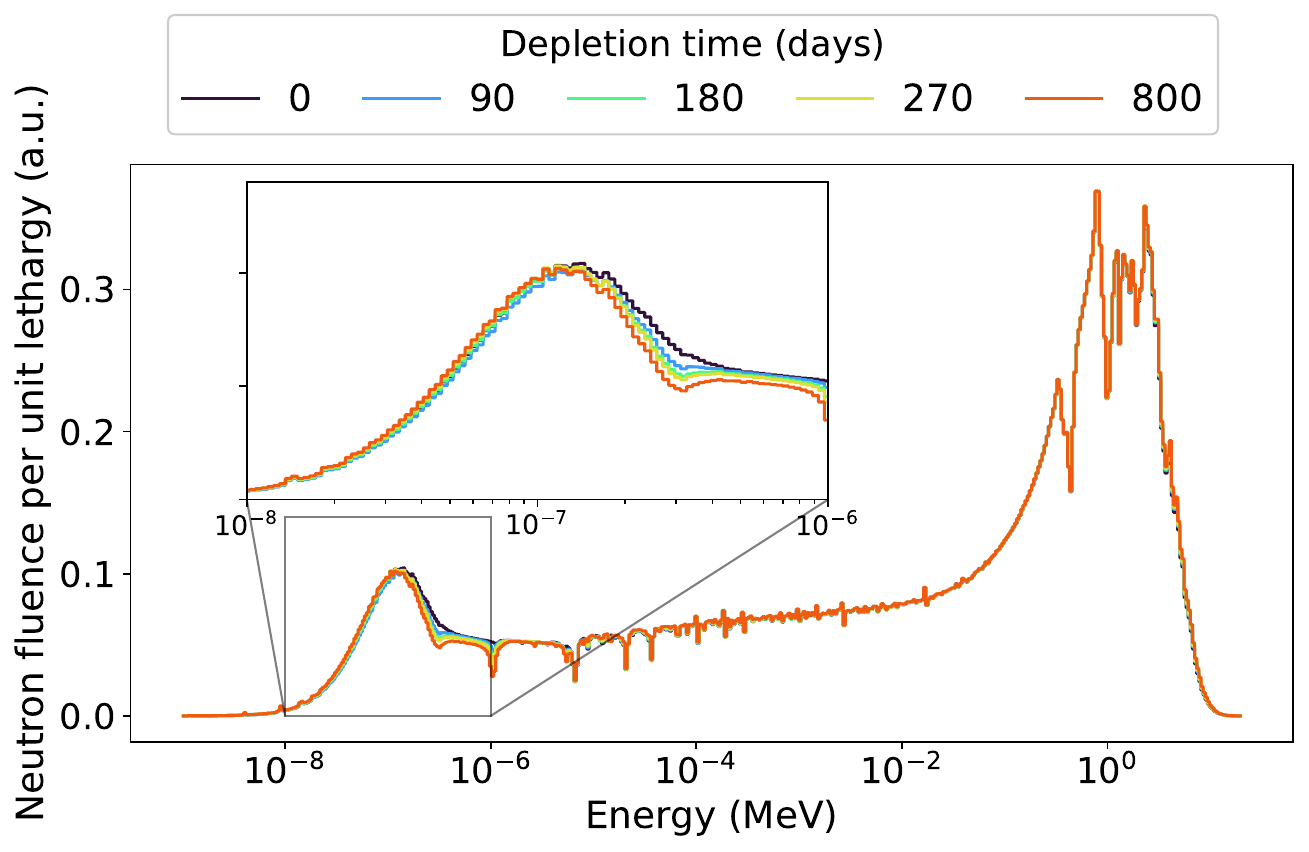}
    \caption[Energy Spectrum]{Energy distribution of neutrons within the SF95 fuel per unit lethargy. Fluence is represented for several depletion steps, from the start of the cycle up to 800 days. Flux behaviour within $10^{-8}$ MeV and $10^{-6}$ MeV is very important in light water reactors, being the region with the highest fission probability from fissile isotopes.}
    \label{fig:FreshFuel_energy}
\end{figure}

In Fig.~\ref{fig:FreshFuel_energy}, we show the neutron energy distribution at several depletion steps, obtained by setting the spatial domain in Eq.~\eqref{eqn:ReactionRate} equal to the entire fuel volume and by dividing the energy range into 500 bins from $E = 0.001$ eV to $E=20$ MeV (with the same width in logarithmic scale\footnote{This subdivision of the energy bins is also called \textit{lethargic scale}.}). In light water reactors, the energy region within $10^{-8}$ MeV and $10^{-6}$ MeV (also called, thermal energy region) plays a crucial role, since the fission interactions are more probable within this range. Depletion as well other design parameters, modify the evolution of the thermal region, having an impact on the reactor performances as well as on the fission fraction evaluation.

In order to map the evolution of nuclide concentrations along the axial direction and compare them with the experimental data, we divided the fuel pins into 30 burnup zones along the z axis. This means that in each volume the fuel burnup is computed starting from the neutron flux evaluated in it.

As introduced in Section \ref{sec:ReactorParameters}, a burnup simulation implies the coupling of two equations. The first one is the Boltzmann equation, which determines the neutron flux by knowing the material composition of the system (i.e., the isotopic concentrations) and is solved using the Monte Carlo routine. The second one is the Bateman equation, solved through a direct numerical method, which provides the time evolution of the materials in the system when the flux distribution at an initial time is known. These two equations are solved alternately during the depletion chain accordingly to a burnup scheme. The time discretization choice (i.e., how many times to solve the Boltzmann equation) is nontrivial.
In this work we divided the time frame into two parts: 
\begin{itemize}
    \item in correspondence with the startup at time $t_0$, we set 5 time steps in an irregular, ascendant order between $t_0$ and $t_0 + 2$ d, namely: $t_0$, $t_0+0.1$, $t_0+0.2$, $t_0+0.5$, $t_0+1$ and $t_0+2$. In this way we take into account the buildup of those fission products that have a large absorption cross section (the so called \textit{poisons}, e.g. the $^{135}\text{Xe}$);
    \item from $t_0 + 2$ d up to the shutdown time $t_d$, we use a uniform time step length of $\Delta t = 50$ d.
\end{itemize}
We run Serpent simulations using an implicit scheme (Stochastic Implicit Euler, SIE) in order to avoid spatial instabilities that occur in 3D burnup calculations~\cite{BurnupMethods}. 
This method, in a similar way to all implicit schemes, solves the differential equation adopting the (unknown) end of step (EOS) flux value. At the beginning of each burnup step, a predictor passage of Forward Euler is applied. Then, several corrector iterations take place. Each iteration adopts the EOS flux from the previous corrector step. After repeating this procedure a fixed number of times, the final estimation of the nuclide density is obtained with the average of the corrector fluxes. More details about the SIE method can be found in~\cite{SerpentSIE,CastagnaImplicit}).
The adopted statistic for each case is 160 active cycles + 100 inactive cycles in the predictor step, with 6$\times10^{5}$ neutrons/cycle and 16 substeps for the implicit scheme.

\begin{figure*}[ht!]
    \centering 
      \includegraphics[width=0.9\textwidth]{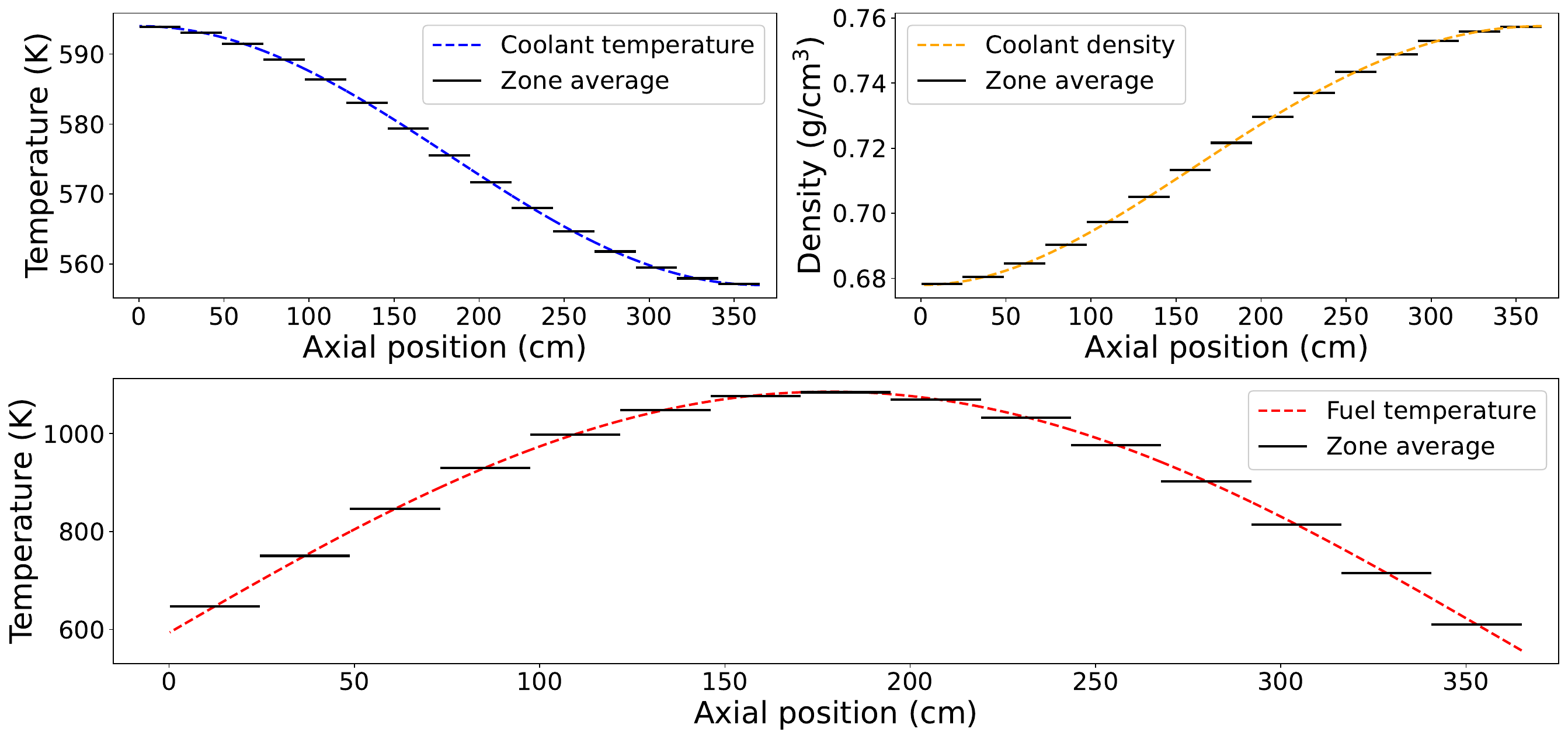}\\
\caption{\label{fig:TemperatureDensity} Description of the temperature and density profiles for the fuel and coolant along the z axis: solid horizontal lines are the values applied to the simulated Serpent regions, while the dashed lines are the continuous values based on design parameters. The zero of the axial position coincide with the top of the rod, accordingly to the Takahama report.}
\end{figure*}

\subsection{Temperature treatment}
\label{sec:temperatureTreatment}
In order to estimate the effect of the temperature field treatment characterizing the fuel assembly, we implemented two different approaches: 
\begin{enumerate}[(i)]
    \item the same temperature $T_{\text{c,f}} = 900$ K is applied to all 30 zones\footnote{The 30 zone divisions apply for all the fuel pins in the assembly.} and water is set at $T_{\text{c,w}} = 575$ K everywhere, with the correspondent density evaluated at p = 155.132 bar of $\rho_{\text{c,w}} = 0.7227$ g/cm$^3$. Average value data are applied coherently with the experiment report \cite{Takahama3}.
    \item the 30 burnup zones are grouped into 15 sub-zones, applying an axial dependent value for both the fuel temperature as well as the coolant temperature (and, consequently, density). Indeed, the coolant density value is evaluated from the combined knowledge of temperature/pressure with the pyXSteam~\cite{pyXSteam} library for Python. The coolant profile is chosen accordingly with the shape reported in the original report \cite{Takahama3}, constraining the inlet/outlet temperature to $T_{in} = 557$ K and $T_{out} = 594$ K. The fuel temperature profile is intended to reflect a realistic distribution, superimposing to the coolant temperature, a cosine-shaped curve, as reported in literature \cite{lamarsh2011introduction}, constraining the average value to be equal to 900 K. Analytical expressions for coolant and temperature axial distributions are reported in Eq.~\eqref{eq:temperatureAxis}
    \begin{align}
    \label{eq:temperatureAxis}
        T_c(z) & = T_{in} + c_1 \left(1+ \sin(c_2 z + c_3) \right) \\ \nonumber
        T_f(z) & = T_c(z) + d_1 \cos(d_2 z + d_3)
    \end{align}
    
    The parameters \{$c_1, c_2, c_3, d_1, d_2, d_3$\} are evaluated constraining inlet/outlet for coolant and average value for fuel. The analytical expressions along with the sub-zones averaged values adopted in the Serpent simulation are represented in Figure~\ref{fig:TemperatureDensity}.

\end{enumerate}

\subsection{Boron treatment}
\label{sec:boronTreatment}
In order to maintain constant power over time, a nuclear plant needs to be in a critical condition, keeping the multiplication factor k$_{\text{eff}}$ equal to 1. To achieve this goal, in light water reactors, a certain amount of boric acid is dissolved into the coolant, such that the neutron balance is modified, increasing the absorption interactions on boron isotopes, and maintaining a balance within the neutron population. The amount of boron dissolved in the system is continuously adjusted over time, due to the combined effect of Gadolinium depletion and fuel burning. The amount of dissolved boric acid is defined by the boron letdown curve. From a physical point of view, boron dissolved in the coolant changes the spectrum thermal region, since its absorption cross section is higher in the thermal region. In this work, two scenarios were considered:

\begin{enumerate}[(i)]
    \item \textit{Automatic boron treatment}: the MC code Serpent has a built-in feature to evaluate the critical boron density at each depletion step such that the k$_{\text{eff}}$ = 1. This feature adds a number of inactive cycles, launched prior to the depletion step, where the critical boron density is evaluated based on the present inventory. Then, the SIE step is performed fixing that amount of boron in the system; 
    \item \textit{Average boron treatment}: the burnup calculation is carried out with a constant boron value, that we set equal to the mean value of the boron letdown curve discussed at the previous point.
\end{enumerate}

\begin{figure}[h]
    \centering
    \includegraphics[width=0.98\linewidth]{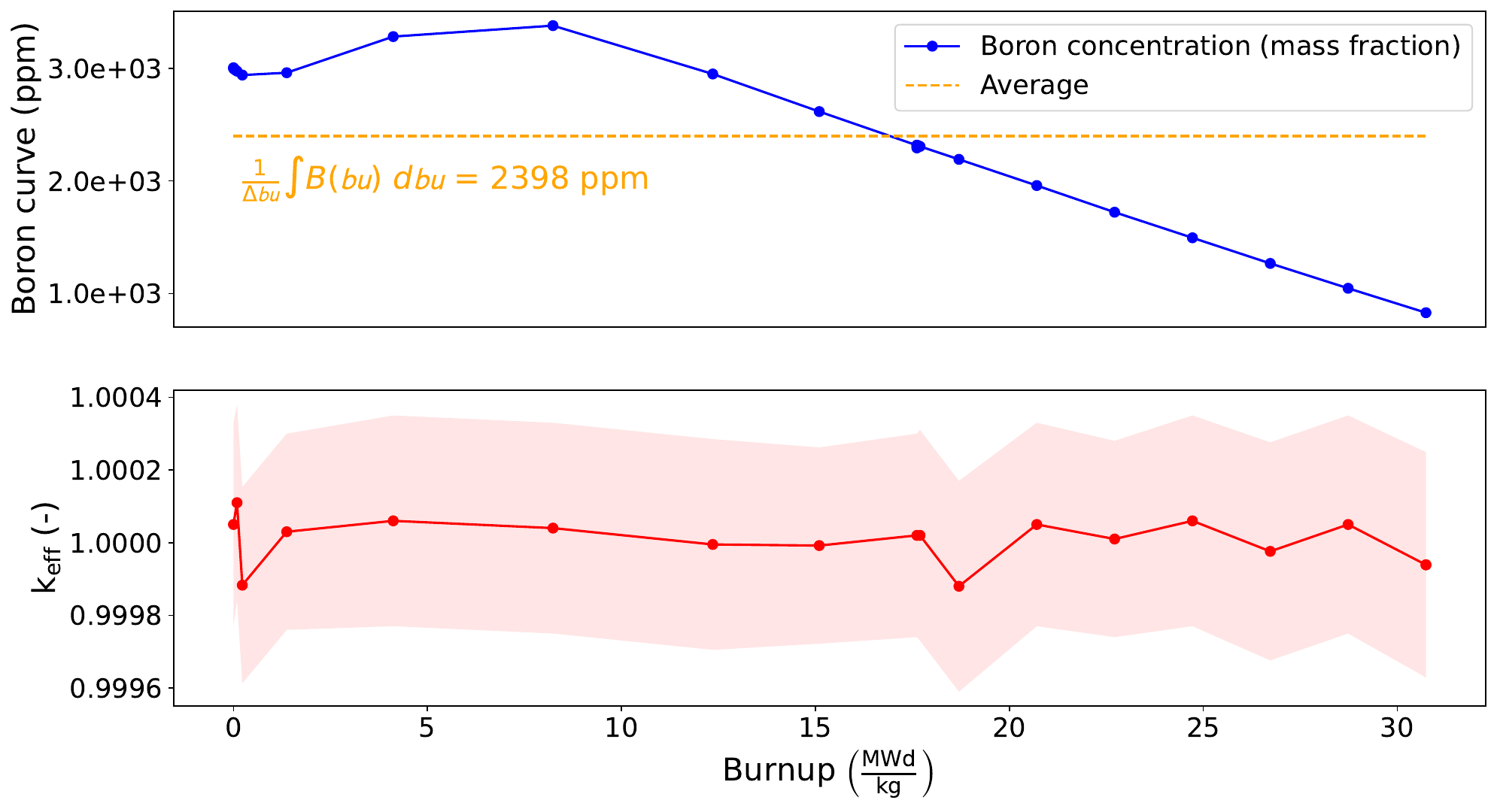}
    \caption[Boron Curve]{Evolution of the boron dissolved within the coolant according to the Serpent simulation: extra number of inactive cycles are added prior to the transport/depletion, evaluating the adjusted boron density (top panel) such that the system is critical (i.e.,  $\text{k}_{\text{eff}}$ is close to one). Bottom panel represents a measure of the system multiplication factor with the associated statistical uncertainty.}
    \label{fig:boronCurve} 
\end{figure}

Both approaches can be seen in Fig.~\ref{fig:boronCurve}, in which the amount of critical boron calculated by Serpent and its average (top panel, blue line and orange dashed line) to maintain the system's reactivity equal to zero (bottom panel) is shown. Since we are simulating a single fuel assembly in an infinite lattice, starting from fresh fuel and without control rod insertion, we obtain a higher value of critical boron concentration with respect to the literature value (\cite{Takahama3}). This analysis is able to check the difference among the results when using different boron treatments (constant boron vs variable boron).
For sake of completeness, it is worth mentioning that an additional scenario is considered to complete the analysis, namely the burnup cycle with the imposition of the boron values reported in original report~\cite{Takahama3}. Even though we acknowledge that this is theoretically the best approach for boron treatment, it is important to highlight that in current reactor operations, the boron value is not a priori known.

\subsection{Simulated scenarios}

\begin{figure*}[ht!]
    \centering 
      \includegraphics[width=0.9\textwidth]{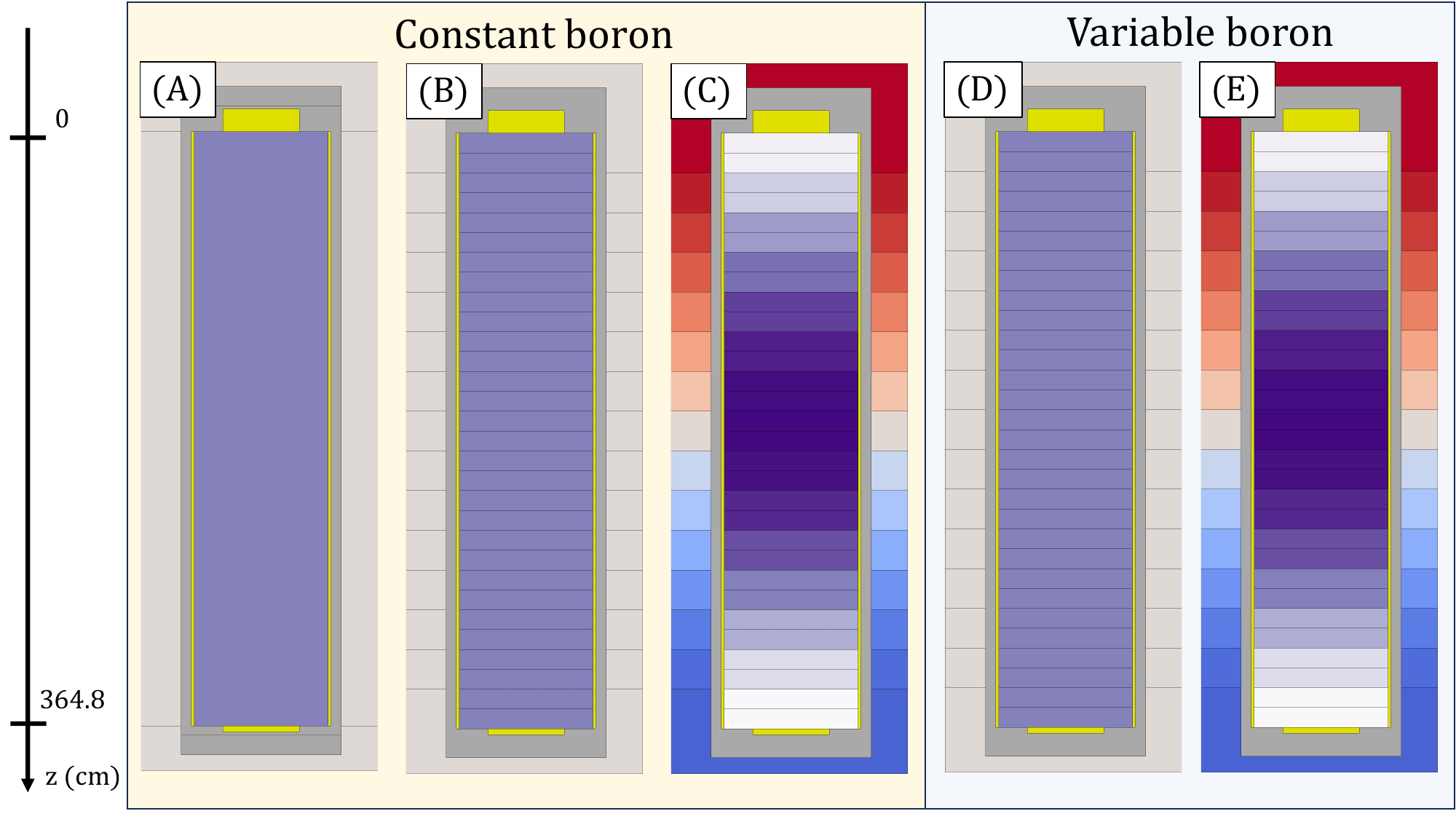}\\
\caption{\label{fig:GeometryModels}Geometries adopted for the description of fuel rods in the Takahama-3 reactor and for the simulation setups. (A) 1 burnup zone at constant temperature for fuel/coolant, in a constant boron environment; (B) 30 axial burnup zones at constant temperature for fuel/coolant in a constant boron environment; (C) 30 axial burnup zones with 15 temperatures for fuel/coolant in a constant boron environment; (D) 30 axial burnup zones at constant temperature for fuel/coolant in a variable boron environment; (E) 30 axial burnup zones with 15 temperatures for fuel/coolant in a variable boron environment; 
values for temperature in cases (C) and (E) is represented in Figure \ref{fig:TemperatureDensity}, and analytical expressions are found in Equation~\eqref{eq:temperatureAxis} . Constant boron values adopted for cases (A), (B) and (C) are evaluated as the integral mean from the boron letdown curve from cases (C) and (E). For consistency with respect to the experimental report the z-axis starts from the top of the fuel active region in the downward direction.}
\end{figure*}

Fig.~\ref{fig:GeometryModels} shows a schematic of the five simulated models, combining the temperature and boron treatments. The first three cases (A,B,C) are set in a constant boron environment. In particular, \begin{itemize}
    \item the first case study (A) is not used for the purpose of the experimental benchmark due to the lack of spatial meshing, but is intended to investigate the role of spatial discretization in the computation of fission fractions;
    \item case (B) exploits a uniform temperature/density treatment, while case (C) imposes the spatially dependent properties;
    \item cases (D) and (E) keep the same features of temperature treatment of (B) and (C) but with the automatic boron strategy. 
\end{itemize}
The additional boron scenario mentioned in Section~\ref{sec:boronTreatment}, only considered for validation purposes, adopts a uniform temperature profile and it will be labelled as (Ref). Temperature and density values applied to this schematic representation, are found in Fig.~\ref{fig:TemperatureDensity}.

\subsection{Results}
\label{sec:results}

This section shows the results of the validation between the Serpent model and the experimental data of the SF95 rod, for all the cases shown in Fig.~\ref{fig:GeometryModels}. The purpose is to identify the modelling differences of a simulation by comparing it with the experimental data to quantify their importance in each case. The SF97 rod was compared with the simplest case (B), having the main purpose of testing the reliability of the model in the presence of an additional fuel cycle. Due to the minor importance for the purpose of this paper, the results of the SF97 rod are shown in Appendix~\ref{app:A}.

The top panels of Figures \ref{fig:Axial_C/E_95} show the comparison between the measured axial concentration of the four fissile nuclides against the Serpent predictions. In the bottom panels we show the \textit{Calculated-to-Experimental} (\textit{C/E}) ratios, i.e. the ratios between the nuclide concentrations predicted by Serpent simulations and those measured experimentally.
The closer the \textit{C/E} ratios are to 1, the more accurate the prediction of nuclide concentration is. 
In these plots we also compare the simulation results obtained with the four models for the temperature/boron field: (B), (C), (D), (E).
As for the Uranium isotopes, both \udtc and \udto appear to be almost insensitive to all setups: the former is reproduced with an average discrepancy from the experimental data of $\sim 12$\%, while the latter remains below the $0.3\%$. The \pudtn turns out to be particularly sensitive to all the effects studied: in the spatial trend of this isotope, it is possible to identify and separate all the simulated case histories. 
In particular, the effect of constant boron vs. automatic boron turns out to be the predominant one: (D) and (E) results are lower (up to 10\%) compared to their constant boron counterparts. The temperature approach, on the other hand, has a secondary effect of tilting the axial concentration, increasing it in the higher temperature zone. 
The effects are most visible on this isotope probably due to the fact that \pudtn is generated during burning, so its entire evolution is dependent on the boundary conditions used. The best case scenario is achieved with Simulation (E), which is able to catch the experimental data trend, with the lowest shift from them.
The \pudqu shows similar trends to \pudtn, but less pronounced, and is better predicted.

In principle, the uncertainty to be associated to \textit{C/E} ratios is given by the propagation of both experimental and simulation uncertainties. 
The uncertainties reported in literature associated to the IDMS measurements are $<0.1\%$ for \udtc and \udto, and $<0.3\%$ for \pudtn and \pudqu.
On the other hand, the uncertainty related to the simulation is not a trivial quantity to retrieve for what concern the results of the depletion scheme. The Serpent code is not able to directly estimate the error associated to the concentrations (\textit{C}) coming from the solution of the Bateman equation, due to its deterministic formulation.  For sake of simplicity, only average values of concentrations are considered, without a correlated uncertainty.

It is worth mentioning that the data in the top locations (SF95-1, SF97-1 and SF97-2) show a worse agreement with the simulation results probably because these samplings were made close to the upper end of the fuel pin. In fact, the estimated uncertainty in the sampling position is declared $<2$~cm~\cite{Suyama2008}. This is a negligible error in the central pin zone, where the flux profile becomes flat as the depletion goes on. On the contrary, at the pin edges where the neutron flux has a significant gradient, a 2~cm shift can introduce a non-negligible systematic error.
Moreover, the neutronics at the axial edges depends on the back-scattering capability of the reflector. Unfortunately, the Takahama-3 reactor design data for reflector and cladding cap length and composition are not detailed in the technical reports. Their dimensions and compositions were hypothesised according to~\cite{Suyama2008}, thus we simulated 2~cm of zircaloy-4 for cladding caps and a 20~cm layer of water. 

For this reason, the following data analysis does not take into account the sampling performed in SF95-1, SF97-1 and SF97-2. Table~\ref{tab:Average_C_E_SF95} reports the average \textit{C/E} for the considered samples (SF95 2-5). In Appendix~\ref{app:A}, the results for SF97 is reported for the samples SF97 3-6 in Table~\ref{tab:Average_C_E_SF97}. A very good agreement is found for \udto whose concentration remains very similar to the fresh fuel value, with \textit{C/E} falling within $\pm0.5\%$. The other three fissile isotopes (\udtc, \pudtn, and \pudqu) exhibit differences depending on the sample site and the simulation case. However, an overall better agreement is found for the (E) case, exhibiting the best \textit{C/E} for almost all sampling sites. By comparing the simulations performed with the  two models of the temperature field (uniform vs axial profile), we observe local effects increasing or decreasing the nuclide concentrations up to 0.7\%. Thus, the temperature field has a relatively small impact if compared with the systematic differences between experimental data and simulations. From the other side, the boron strategy is responsible for an higher difference, increasing the prediction performances up to 10\% for the \pudtn.
In these tables we also report, for comparison, the results published in literature obtained with other simulation tools (evaluating the average \textit{C/E} avoiding the values at the rod extremes), namely: SWAT and ORIGEN 2~\cite{PIE_ORNL}, SCALE 5.1~\cite{SCALE_PIE_ORNL} and SCALE 4.4 and HELIOS~\cite{Takahama3}. 
A similar analysis was previously conducted by \cite{Takahama_DRAGON_MURE}, comparing a deterministic code (DRAGON) and a Monte Carlo (MURE). The results obtained in this work with Serpent simulations are characterized by similar \textit{C/E} values when compared with those obtained in other analyses. For what concerns the excluded data (i.e. SF95-1, and SF97 1-2), results from literature agree with our analysis, with differences with respect to unity up to 35\%.

\begin{figure*}[htbp]
\centering
\includegraphics[width=1\linewidth]{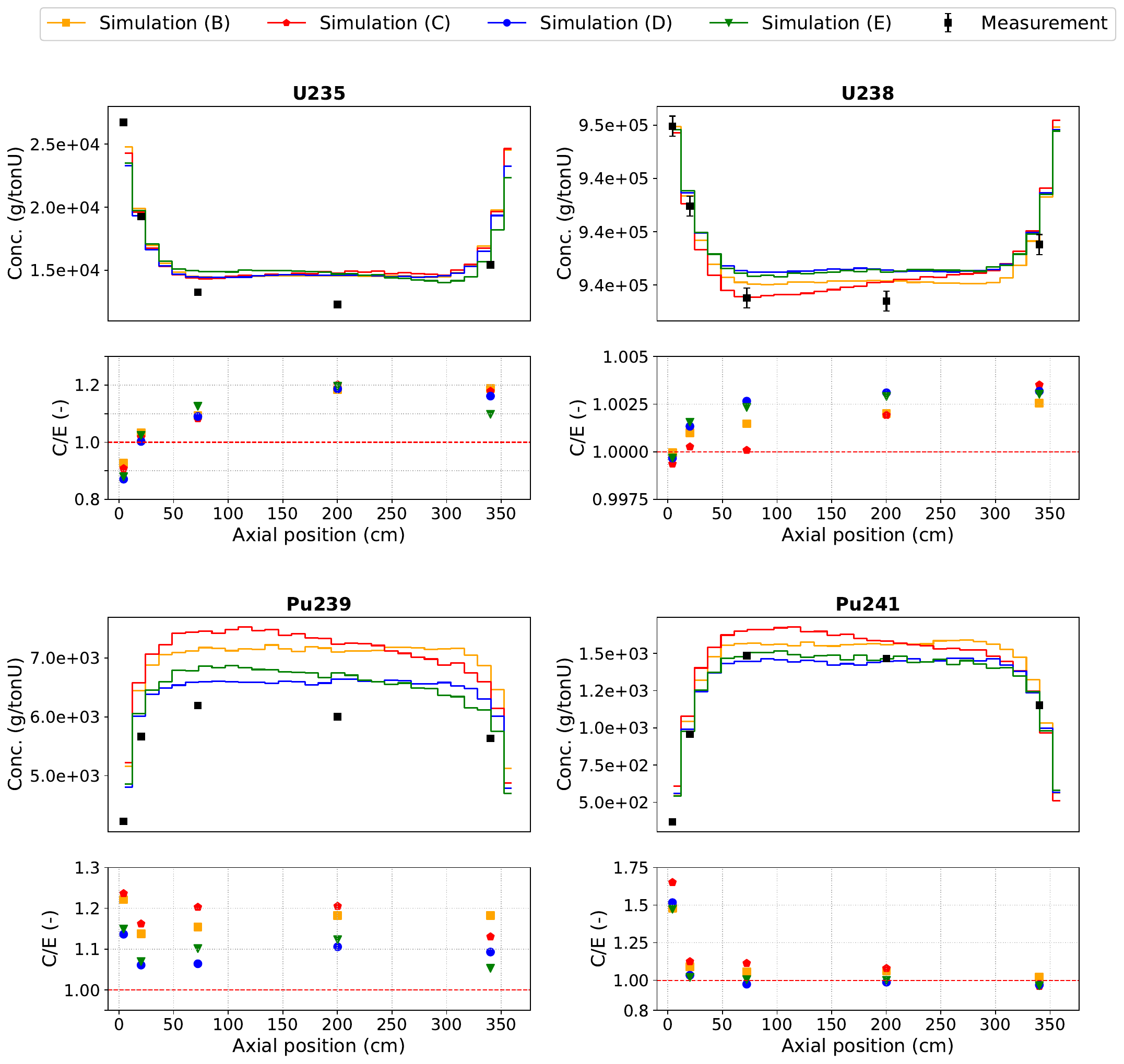}
\caption{\textit{Top panels}: measured axial concentrations of the four fissile isotopes (black dots) in the SF95 pin, compared with simulation results obtained with the design setups B, C, D and E, previously defined. \textit{Bottom panels}: \textit{C/E} ratio of calculated (C) concentrations versus experimentally measured (E) ones for SF95 rod.}
\label{fig:Axial_C/E_95}
\end{figure*}

\begin{table}[h]
\centering 
    \renewcommand{\arraystretch}{1.5}
    \begin{tabular}{c | c | c c c c}
    \hline
     \textbf{Rod} & \textbf{Code} & \textbf{$^{235}$U}   & \textbf{$^{238}$U} & \textbf{$^{239}$Pu} & \textbf{$^{241}$Pu}  \\
      \hline\hline
      \multirow{9}{*}{\textbf{SF95}} & Serpent (B)  & 1.124 & 1.002 & 1.164 & 1.058 \\
     & Serpent (C)  & 1.120 & 1.001 & 1.175 & 1.07 \\
     & Serpent (D)  & 1.110 & 1.003 & 1.081 & 0.991 \\
     & Serpent (E)  & 1.111 & 1.002 & 1.087 & 0.997 \\
     & Serpent (Ref)  & 1.083 & 1.003 & 1.042 & 0.955 \\      
     & SWAT  & 1.015 & 1.0 & 1.009  & 0.959 \\
     & ORIGEN 2.1 & 1.016 & 1.0 & 1.008 & 0.976 \\ 
     & HELIOS & 1.008 & 1.0 & 1.026 & 1.01 \\ 
     & SCALE 4.4 & 0.985 & 1.0 & 0.976 & 0.95 \\ 
     & SCALE 5.1 & 1.028 & 1.002 & 1.097 & 1.017 \\ 
     \hline
    \end{tabular}
    \\[10pt]
    \caption{Average Serpent \textit{C/E} for chosen samples (SF95 2-5) in the selected cases B, C, D, E and Ref, compared with the literature results obtained with different simulation codes.}
    \label{tab:Average_C_E_SF95}
\end{table}

\section{Fission fractions and \antinu spectra results}
\label{sec:antinu_spectra}
In the framework of reactor \antinu spectrum prediction, the estimation of the fission fractions of fissile nuclides is more relevant than their concentrations.
In this section, first we present the time evolution of fission fractions obtained with different levels of approximation (Sect.~\ref{sec:FF}). Then, we show the non-oscillated \antinu spectrum, at different fission fractions values (Sect.~\ref{sec:NonOscillatedSpectrum}).

\subsection{Fission Fractions}
\label{sec:FF}
\begin{figure*}
     \centering
    \includegraphics[width=\linewidth]{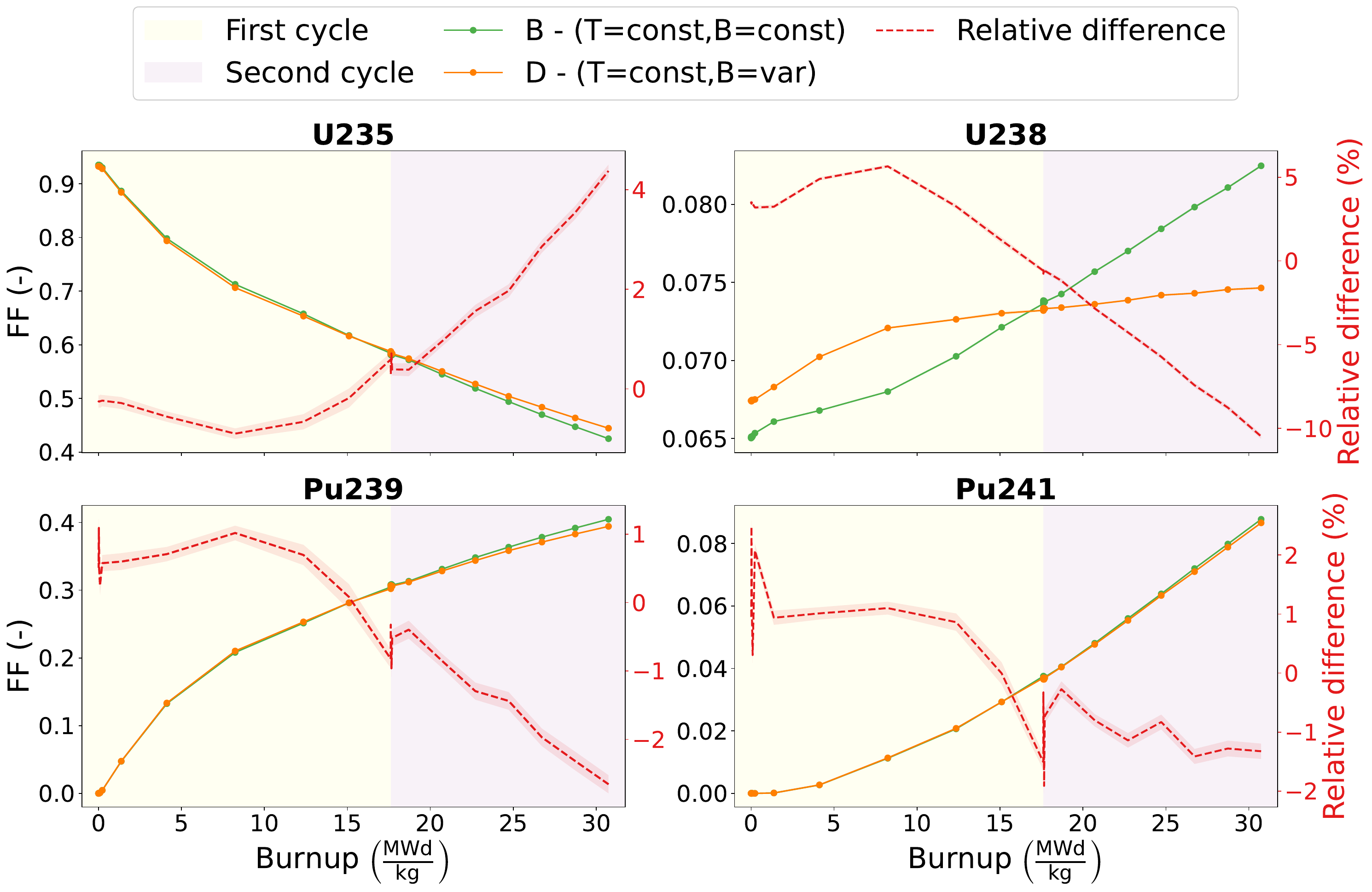}
    \caption{Comparison between fission fraction in simulation B (green line) and simulation D (orange line) and their relative difference (red dashed line). The effect of the boron treatment (automatic against average) on the fission fraction calculation shows a trend for every fission fraction. Fission events in \udto are the one that suffer most from the boron level in the system, completely changing the slope of the time evolution, with a maximum difference of -10 \%.}
    \label{fig:FF_95_BD}
\end{figure*}

\begin{figure*}
     \centering
    \includegraphics[width=\linewidth]{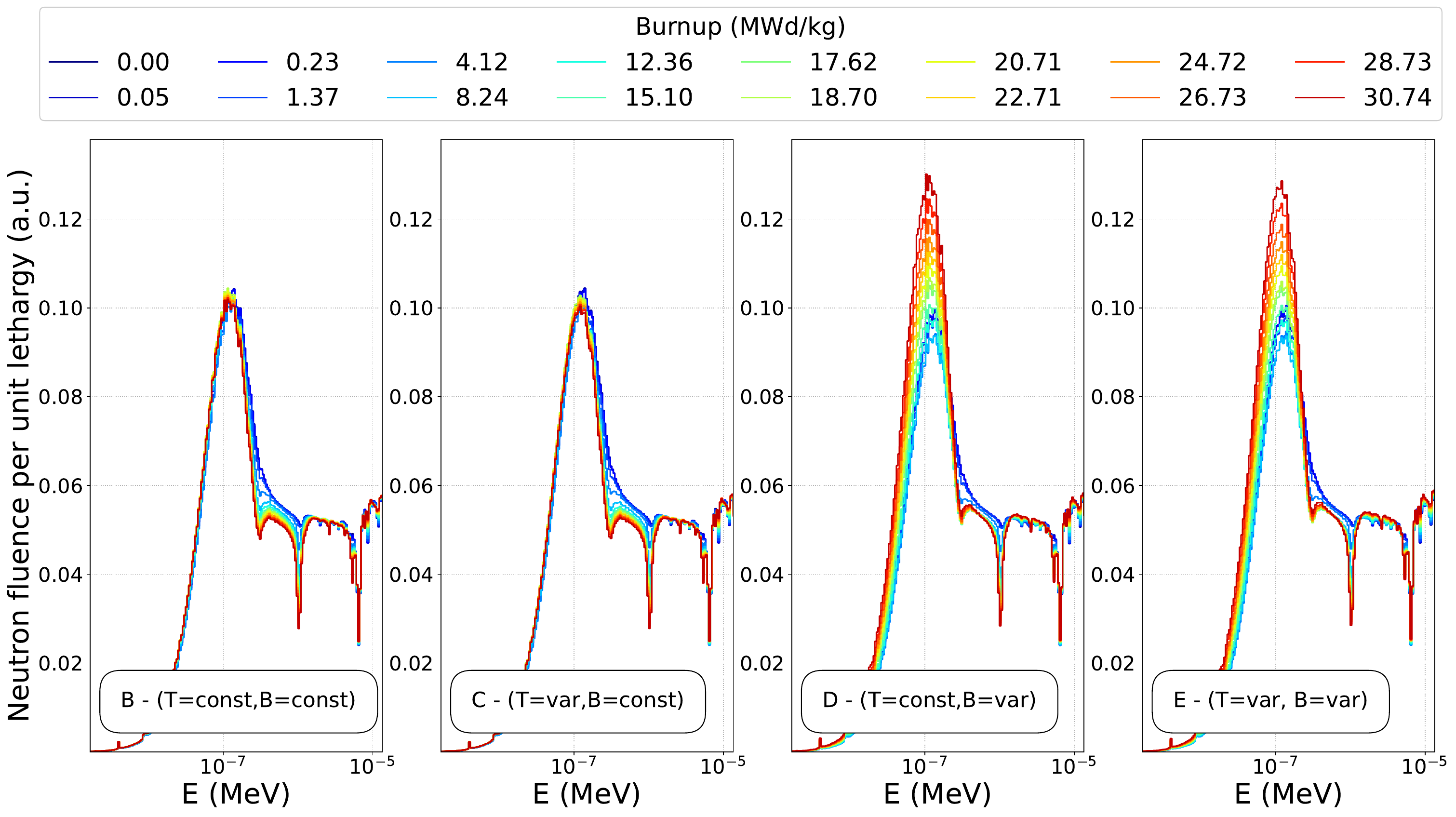}
    \caption{Neutron fluence in the thermal energy range for the various simulation cases (B to E) during the burnup evolution (from 0.0 MWd/kg up to 30.74 MWd/kg). 
    Simulation setups where automatic boron is activated (i.e., D and E) shows an higher change in the neutron energy distribution in the range where fission is more probable (for \udtc, \pudtn and \pudqu).} 
    \label{fig:PhiE_tot}
\end{figure*}

Among the possible combinations of cases simulated according to the scheme in Fig.~\ref{fig:GeometryModels}, we want to give importance to the principal effect, which was also shown in the validation reported in the previous section, namely, the choice of the boron strategy. Figure~\ref{fig:FF_95_BD} shows the trend of fission fractions in the standard fuel in the case with uniform temperature and constant boron (B) versus uniform temperature and automatic boron (D), with their relative difference. With fresh fuel, the contribution to the fission events comes from \udtc ($\sim$ 94\%) and \udto ($\sim$ 6\%). As the fuel depletion proceeds, two major effects can be appreciated: \udtc is consumed and Pu nuclides are produced from \udto neutron captures. 
Therefore, with the increase in the burnup level, the relative importance to the total fission events of \udtc decreases, while the \pudtn and \pudqu contribution increases. 
The different background colors represent the two simulated cycles, each characterized by a different nominal power. For the presented case, the (B)-(D) comparison shows a clear dependence on the burnup level: the relative difference between the fraction increases as a function of the burnup level. In particular, the fraction of the \udto is the one most sensible to the boron treatment. To explain this phenomenon, we should look at what happens in the thermal region of the neutron spectrum in the different cases: as depicted in Fig.~\ref{fig:PhiE_tot}, the change in the boron level consist in a very large modification of the Maxwellian peak: when the boron concentrarion is lower, the peak is higher and viceversa. This fact influences how many fissions occur in fissile isotopes (\udtc,  \pudtn or \pudqu) and how many in the fissionable (\udto).

Table~\ref{tab:CycleAveragedFF} report the cycle-averaged fission fractions for all the presented scenarios, for both cycle 5 and cycle 6. Again, the highest differences can be appreciated comparing automatic boron cases against constant boron cases. Here, is reported the relative difference in the (B)-(D) comparison for both cycles:
\begin{itemize}
    \item (\udtc) $-0.5\%$ for cycle 5 and $+2.0\%$ for cycle 6;
    \item (\udto) $+3.5\%$ for cycle 5 and $-5.4\%$ for cycle 6;
    \item (\pudtn) $+0.5\%$ for cycle 5 and $-1.5\%$ for cycle 6;
    \item (\pudqu) $+0.2\%$ for cycle 5 and $-1.1\%$ for cycle 6.
\end{itemize}

Another interesting difference can be appreciated comparing the cycle averaged fractions of case (A) against the others: the effect of the number of burnup zones is much more pronounced in the first cycle with respect to the second. 
The precise comparison of the other cases (temperature effect, mesh effect) will be faced in Appendix~\ref{app:B}. For what concern the boron simulation strategy, we can state that its modelling in MC simulations has a non-negligible effect in the estimation of fission fractions, especially the \udto one, where a non correct approach may lead to a fission fraction overestimation/underestimation, depending on the cycle. 

\begin{table}[ht]
\centering 
    \renewcommand{\arraystretch}{1.2}
    \begin{tabular}{c p{1cm} p{1cm} p{1cm}  p{1cm} p{1cm}}
    \hline
       & \textbf{Case} & \textbf{\udtc} & \textbf{\udto} & \textbf{\pudtn} & \textbf{\pudqu} \\
      \hline\hline
      \multirow{5}{*}{Cycle 5} & (A)  &  0.734 & 0.068  & 0.185 & 0.012 \\
                               & (B)  &  0.724 & 0.069  & 0.193 & 0.014 \\
                               & (C)  &  0.725 & 0.069  & 0.192 & 0.014 \\
                               & (D)  &  0.721 & 0.071  & 0.194 & 0.014 \\
                               & (E)  &  0.720 & 0.072  & 0.194 & 0.014 \\
      \hline
      \multirow{5}{*}{Cycle 6} & (A)  & 0.502  & 0.078  & 0.359 & 0.061 \\
                               & (B)  & 0.502  & 0.078  & 0.358 & 0.062 \\
                               & (C)  & 0.500  & 0.079  & 0.359 & 0.062 \\
                               & (D)  & 0.512  & 0.074  & 0.352 & 0.061 \\
                               & (E)  & 0.510  & 0.075  & 0.354 & 0.061 \\
      \hline
    \end{tabular}
    \caption[Averaged FF]{Cycle-averaged fission fraction for the simulated scenarios.}
    \label{tab:CycleAveragedFF}
\end{table}

\subsection{Non oscillated \antinu spectra results}
\label{sec:NonOscillatedSpectrum}

\begin{figure*}
    \centering
    \includegraphics[width=\linewidth]{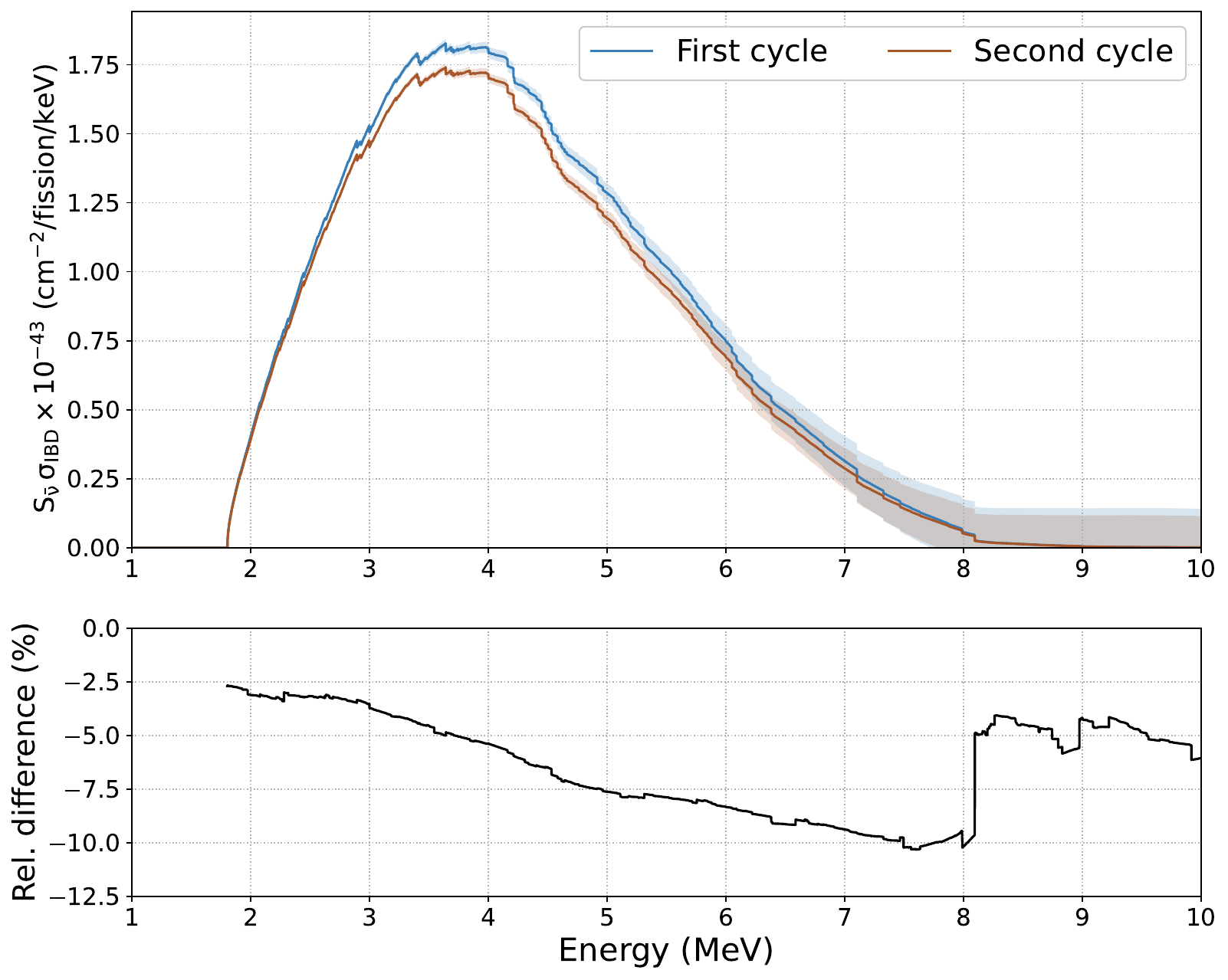}
    \caption{\textit{Top}: \antinu spectra multiplied by the IBD cross section computed with the fission fractions corresponding to the cycle averages value in the simulation (B). \textit{Bottom}: Relative differences of the two spectra.}
    \label{Fig:antinu_FF}
\end{figure*}

\begin{figure*}
    \centering
    \includegraphics[width=\linewidth]{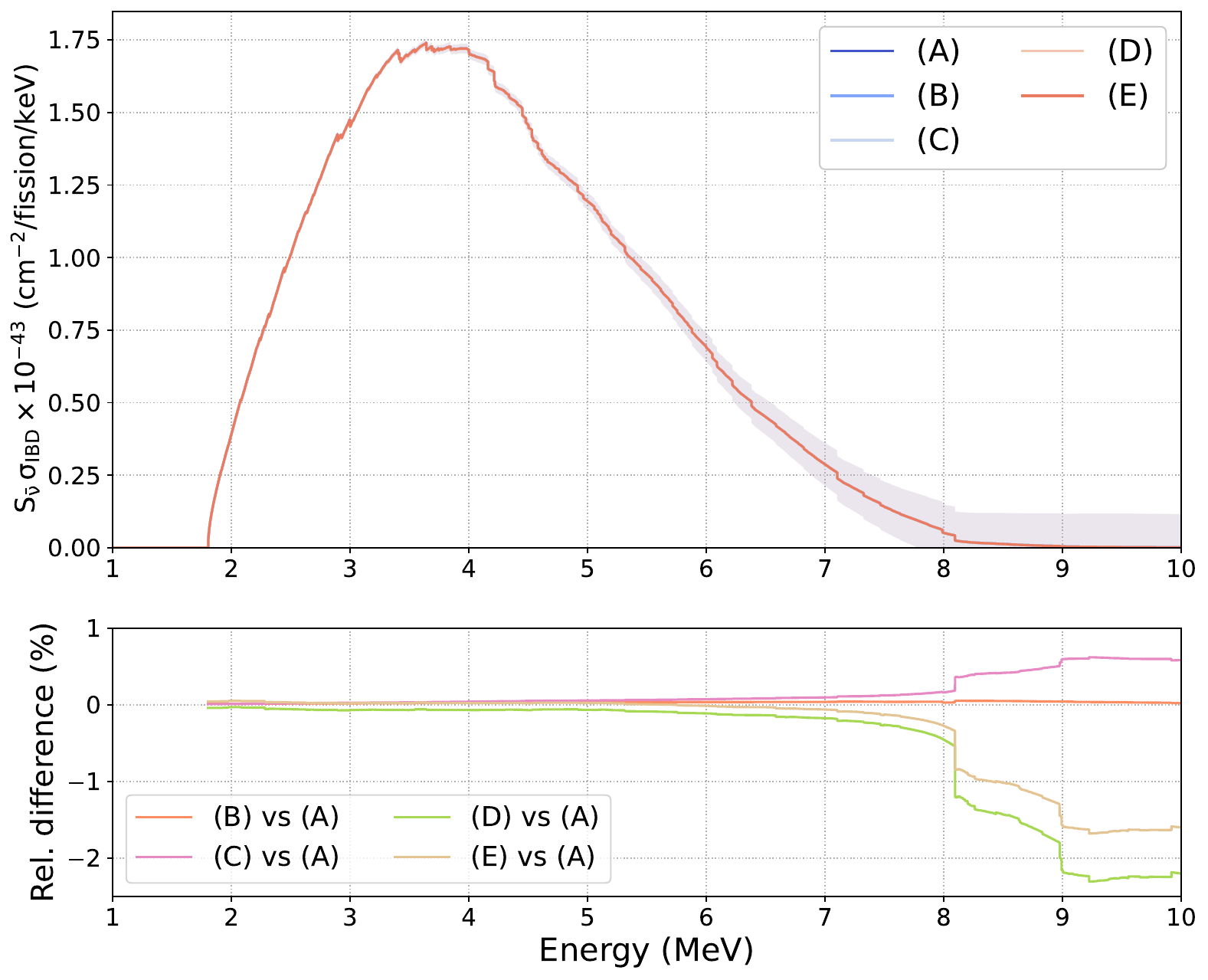}
    \caption{\textit{Top}: \antinu spectra multiplied by the IBD cross section computed with the last averaged cycle fission fractions for the cases considered. \textit{Bottom}: Relative differences of the spectra with respect to the (A) case.}
    \label{Fig:antinu_FF_model_comparison}
\end{figure*}

We finally present the equilibrium \antinu spectra obtained through the summation method by combining the cumulative fission yield (CFY) data published in the Live Chart of Nuclides by IAEA~\cite{IAEA_livechart} and the \antinu spectra of 468 \bm decays obtained through BetaShape by setting an energy binning of 1~keV.
We use thermal CFY for \udtc, \pudtn and \pudqu, which are mostly fissioned by thermal neutrons, and fast CFY for \udto which can be fissioned with not negligible probability by fast neutrons only.
Then, the total reactor \antinu spectrum (Equation \eqref{eqn:SpectrumEquationI}) is evaluated with the fission fractions estimated with Serpent simulations. Since the reactor \antinu will be detected through the IBD reaction, we multiply the \antinu spectra by the IBD cross section~\cite{STRUMIA200342} so as to obtain the shape of the detectable non oscillated spectrum.
We want to highlight that this analysis is meant to compare the propagation of the reactor model differences on the reactor antineutrino spectrum. Other reactor antineutrino calculation methods as well as comparison with experimental data is out of the scope of the analysis and will be presented in future works.

In Fig.~\ref{Fig:antinu_FF}, we show the \antinu spectra obtained by using the cycle-averaged fission fractions estimated by Serpent for the Takahama-3 assembly, highlighting the differences between the emission spectrum during the first cycle and the second one. As shown in the top pad, the \antinu flux intensity decreases with time. This is because the average number of \antinu emitted after \pudtn fission is lower with respect to that of \udtc. The depicted error band for each spectrum propagates the uncertainties of the following variables, namely Q-values, branching ratios and fission yields. For sake of simplicity, the uncertainties are treated as independent. The bottom pad shows the relative difference calculated with respect to the fresh fuel \antinu flux is of the order of -5\% around the maximum of the spectrum at 4~MeV, and is up to $\sim -10\%$ in the high energy range. This result reflects the extreme case where the reactor starts from the fresh condition, without any refuelling strategy: in a more realistic scenario, this effects is averaged out considering fuel at different depletion stages.
Nevertheless, in the framework of the JUNO experiment, it will be very important to correctly sum up the non-oscillated spectra taking into account the fission fraction evolution of each of the eight reactors during the data taking. 

Finally, in Fig.~\ref{Fig:antinu_FF_model_comparison} we show the impact of the different simulated scenarios discussed in Sect.~\ref{sec:FF} on the non oscillated \antinu spectrum.
In particular, the spectra shown in the top pad were computed by using the cycle-averaged fission fractions for the second cycle, obtained from the simulations (A), (B), (C), (D) and (E). 
In this case, the differences are much less pronounced, remaining  below 0.2\% up to 7~MeV and in any case below 2\% at higher energies. This can be addressed to the fact that the major differences are related to the \udto fraction, which has a minor impact on the overall system.

\section{Conclusions}
\label{sec:conclusions}
In this paper we presented a methodology for the prediction of the non-oscillated reactor \antinu spectrum. 
This is intended to serve the scientific community in future reactor \antinu experimental campaigns (e.g., JUNO and TAO) that will analyze the mass ordering problem with an unprecedented level of precision. 
We coupled the Monte Carlo code Serpent along with the \antinu spectrum calculation software BetaShape. 
The Serpent simulations allow the calculation of fission fraction evolution over time, leveraging data encompassing reactor geometry, power history, and fresh fuel inventory. Subsequently, by integrating BetaShape alongside nuclear databases, we construct the antineutrino spectra emanating from the four primary fissile isotopes with the summation method -- \udtc, \udto, \pudtn, \pudqu\ -- thereby enabling predictions of the total reactor antineutrino spectra across various burnup levels. \\
In this paper we focused our analysis on the fuel burnup, aiming at exploring the accuracy in the evaluation of fission fractions through MC simulations. For this purpose, we used the spent fuel concentrations of the Takahama-3 PWR as experimental benchmark for the Serpent simulations. We found that Serpent simulation run with either uniform temperatures in the fuel/moderator or imposing different temperature zones, are both in a reasonably good agreement with the experimental data. Importantly, our simulations highlight that the temperature field introduces minor discrepancies both in the fuel depletion calculations  as well as when considering the fission fractions. We also focused on the boron strategy, comparing the usage of an automatic critical boron calculator and the approach with an constant boron density. When the automatic critical boron is present, the thermal neutron flux is highly influenced by this field, leading to a not-negligible difference in the concentration estimation (\textit{C/E} improved by 10\%) as well as in the fission fraction evaluation (relative difference up to 5\% ).

Looking ahead, our future plans encompass harnessing this analytical tool to further investigate additional sources of systematic uncertainty (such as cross-section or fission yield data uncertainties) and their subsequent impact on the antineutrino spectrum.
Future developments will be devoted to relax the equilibrium hypothesis on the fission products, highlighting the importance that this amount of nuclide have on the spectrum. Also, a more detailed analysis of the reactor configuration of the Taishan and Yangjiang plant will be addressed, along with the comparison with other available experimental spectra from near-detector experiments.

\section{Data availability}
The authors confirm that the data supporting the findings of this study are available within the article.

\bibliography{0_main}  
\bibliographystyle{spphys} 

\newpage

\appendix 
\renewcommand{\thesection}{\Alph{section}} 

\section*{Appendix A}
\section{SF97 validation} \label{app:A} 

This appendix shows results for the SF97 rod validation, being analyzed in the simplest case, considering a uniform temperature for fuel and coolant, and a constant boron approach (i.e., simulation (B)).

\begin{figure}[htbp]
\centering
\includegraphics[width=1\linewidth]{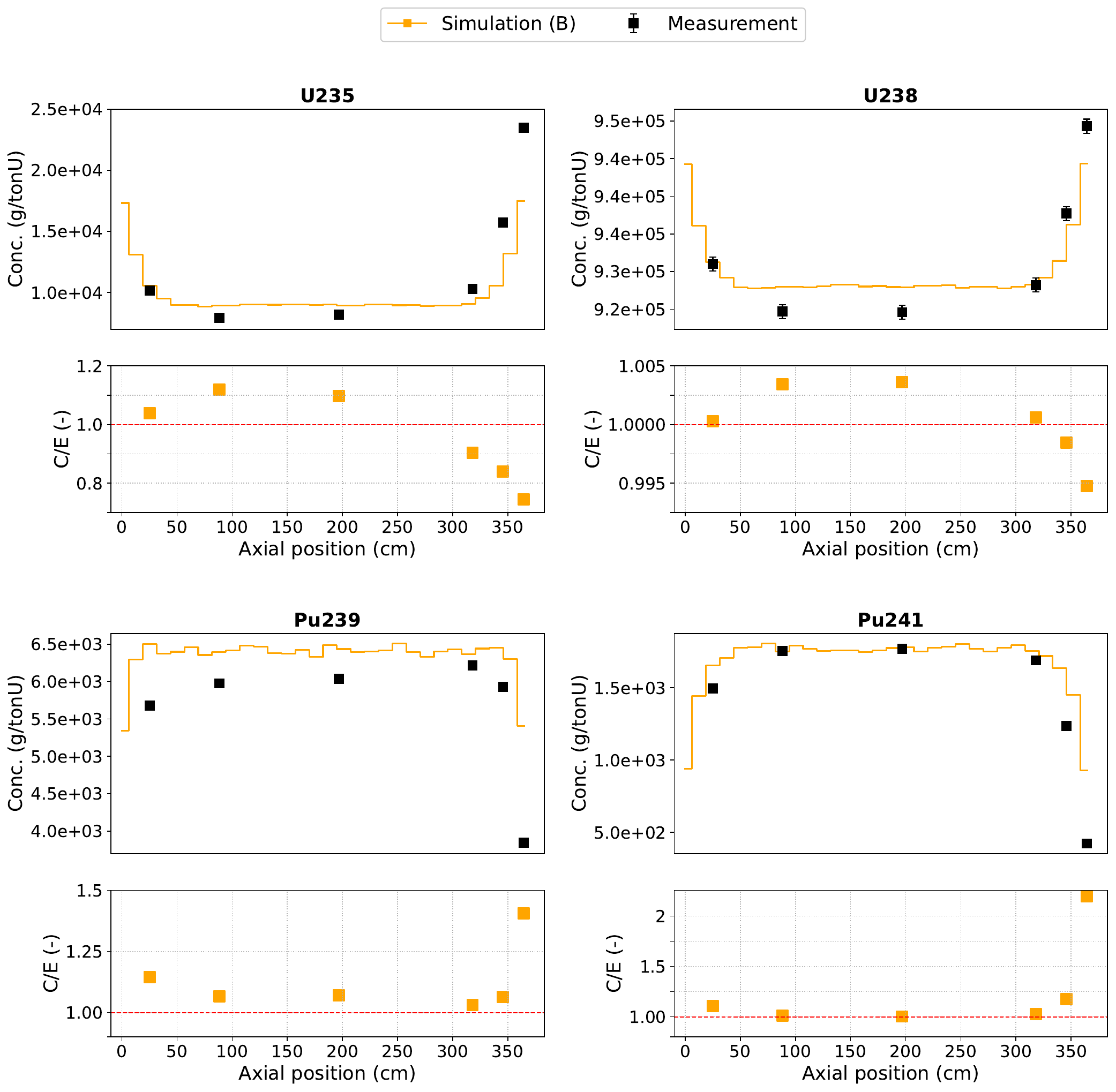}%
\caption{\textit{Top panels}: measured axial concentrations of the four fissile isotopes (black dots) in the SF97 pin, compared with simulation results obtained with uniform temperature field and average boron approach. \textit{Bottom panels}: \textit{C/E} ratio of calculated (\textit{C}) concentrations versus experimentally measured (\textit{E}) ones for SF97 rod.}
\label{fig:Axial_C/E_97}
\end{figure}

The top panels of Figures \ref{fig:Axial_C/E_97} show the comparison between the measured axial concentration of the four fissile nuclides against the Serpent predictions, in the scenario (B). Bottom panels show the \textit{C/E} ratio, giving a quantitavive measure of how the simulation is close to the measurement. Overall, the comparison shows a good agreement, despite the first sample SF97-1, which is overestimated/underestimated up to a factor 2 for $^{241}$Pu. Problems about comparison with the first two samples are reported in Sect.~\ref{sec:results}. Finally, Table~\ref{tab:Average_C_E_SF97} shows the average \textit{C/E} analysis compared with other literature references.

\begin{table}[h]
\centering 
    \renewcommand{\arraystretch}{1.5}
    \begin{tabular}{c | c | c c c c}
    \hline
     \textbf{Rod} & \textbf{Code} & \textbf{$^{235}$U}   & \textbf{$^{238}$U} & \textbf{$^{239}$Pu} & \textbf{$^{241}$Pu}  \\
      \hline\hline
      
      \multirow{6}{*}{\textbf{SF97}} & Serpent (B)  & 1.040 & 1.002 & 1.078 & 1.038 \\
     & SWAT  & 1.015 & 1.0 & 1.031 & 0.986 \\
     & ORIGEN 2.1 & 1.032 & 1.0 & 1.050 & 1.019 \\ 
     & HELIOS & 1.01 & 1.0 & 1.03 & 1.025 \\ 
     & SCALE 4.4 & 0.97 & 1.0 & 0.993& 0.975 \\ 
     & SCALE 5.1 & 1.02 & 0.999 & 1.048 & 0.975 \\
      \hline
    \end{tabular}
    \\[10pt]
    \caption[Average \textit{C/E} for Takahama-3 experiment]{Average Serpent \textit{C/E} for chosen samples (SF97 3-6) in the uniform temperature case and average boron (B), compared with the literature results obtained with different simulation codes.}
    \label{tab:Average_C_E_SF97}
\end{table}

\section*{Appendix B}
\section{Other Fission fractions} \label{app:B} 

Section~\ref{sec:FF} showed the comparison in fission fraction concerning the boron approach. This appendix aims at comparing the other important scenarios: one burnup zone at constant properties (A) and 30 burnup zones at variable temperature and constant boron (C).

\begin{figure}
     \centering
    \includegraphics[width=\linewidth]{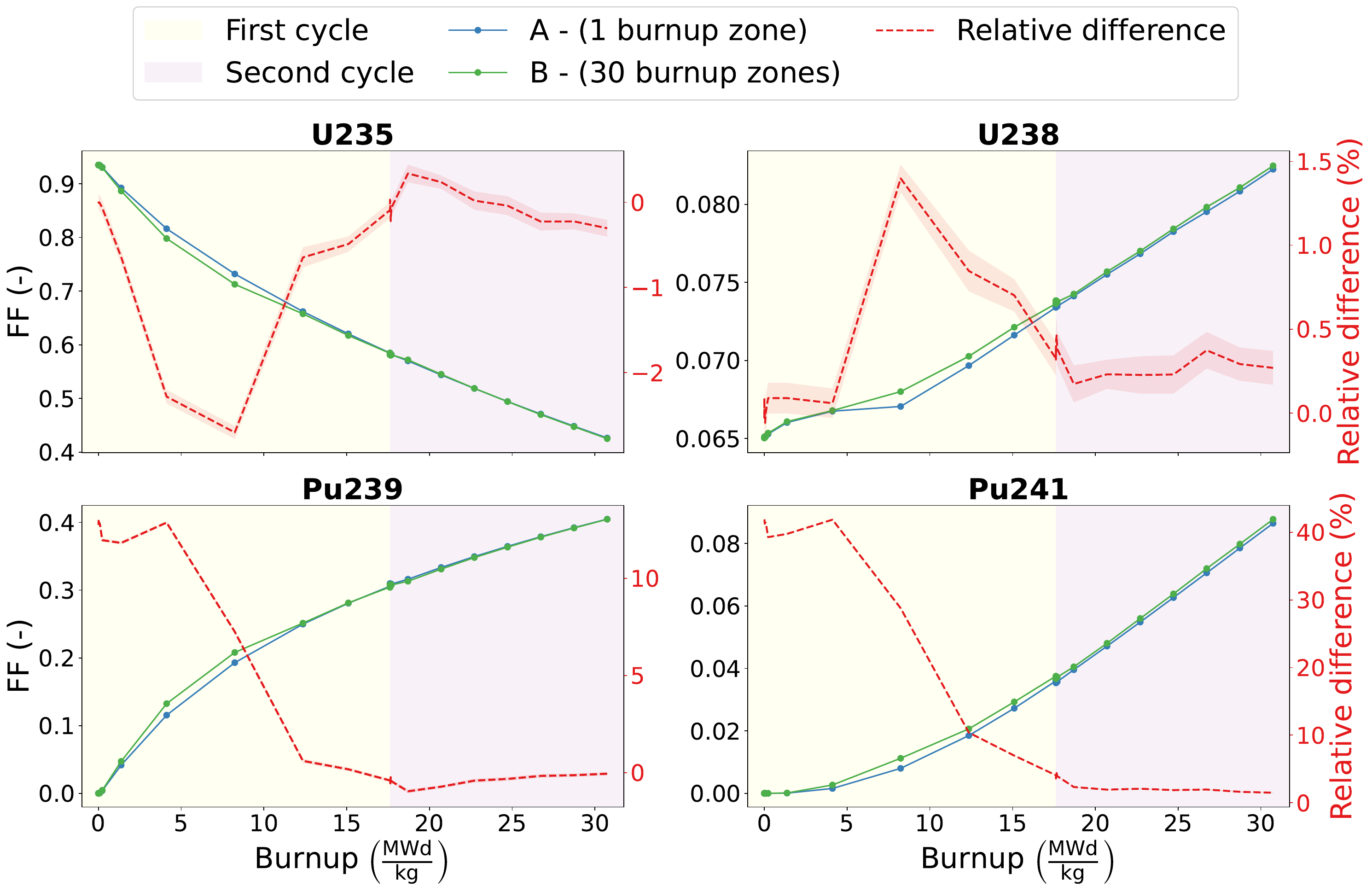}
    \caption{Comparison between fission fraction in simulation A (blue line) and simulation B (green line) and their relative difference (red dashed line). The effect of spatial discretization for the burnup zones on the fission fraction calculation has a maximum in the first burnup cycle, and then tends to a uniform and negligible value in the second cycle.}
    \label{fig:FF_95_AB}
\end{figure}

The burnup discretization effect is showed in Figure~\ref{fig:FF_95_AB}: despite an initial discrepancy, around the burnup level where the gadolinium is depleted, the fission fraction difference tend to zero in the second cycle. From a simulation point of view, this helps in the simulation of large systems, where a fine discretization is not suitable.

\begin{figure}
     \centering
    \includegraphics[width=\linewidth]{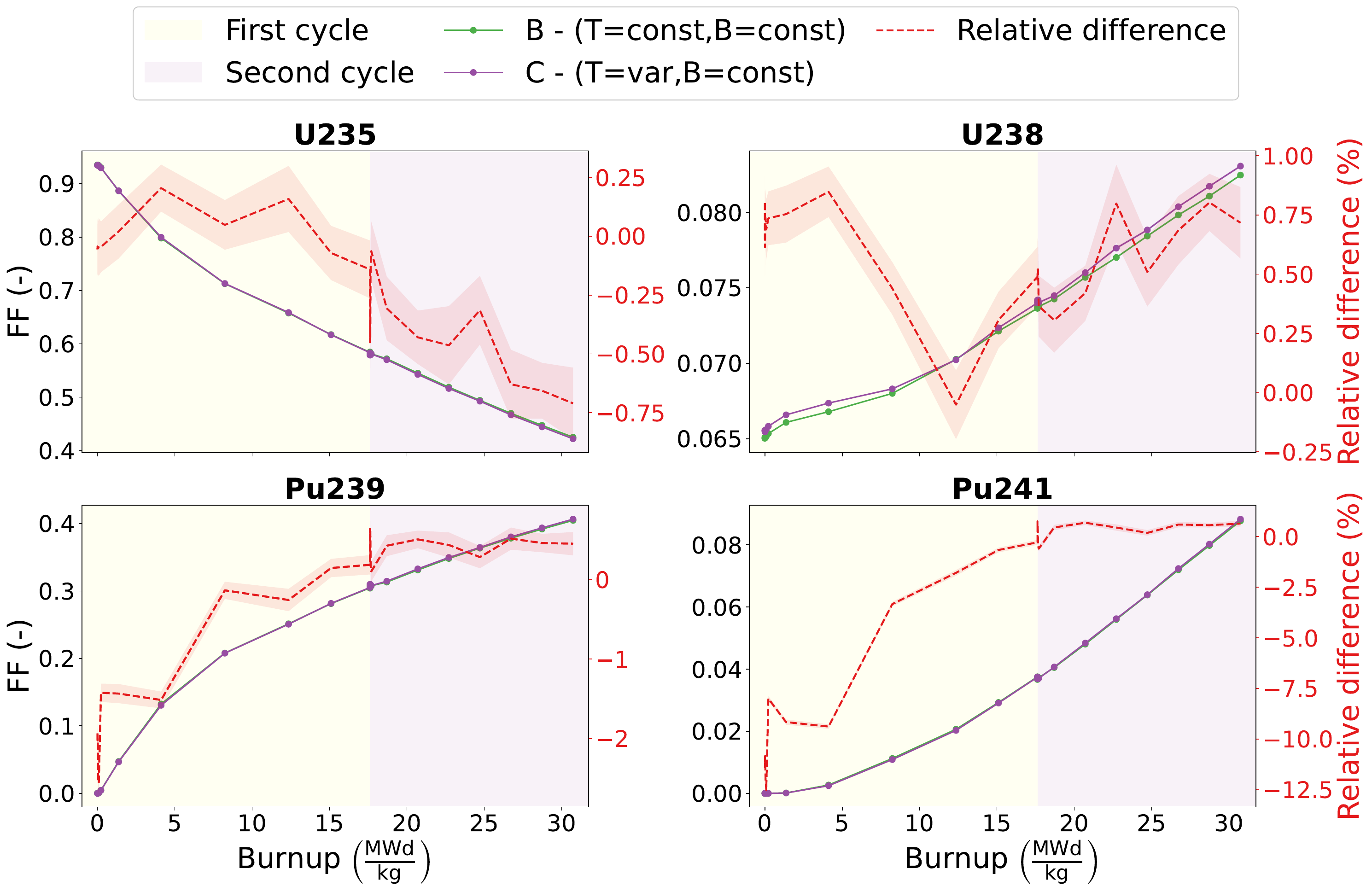}
    \caption{Comparison between fission fraction in simulation B (green line) and simulation C (purple line) and their relative difference (red dashed line). The effect of spatial dependent temperature approach against the average treatment on the fission fraction calculation shows a negligible trend for all the isotopes, with a balance between \udtc (-0.75 \%) and \udto (+0.75 \%).}
    \label{fig:FF_95_BC}
\end{figure}

The impact of the temperature field approach is shown in Figure~\ref{fig:FF_95_BC}: this comparison shows a slight dependence on burnup, decreasing the importance of \udtc and increasing the \udto, laying always below the 1\%. Plutonium fraction effects tend to zero in the second cycle, being unsensible to the temperature approach adopted. Also this effect, give useful information from a simulation point of view, allowing to treat the temperature as an average quantity, helping the description of large systems. Within the JUNO/TAO framework, such difference can be considered negligible.

\begin{figure}
     \centering
    \includegraphics[width=\linewidth]{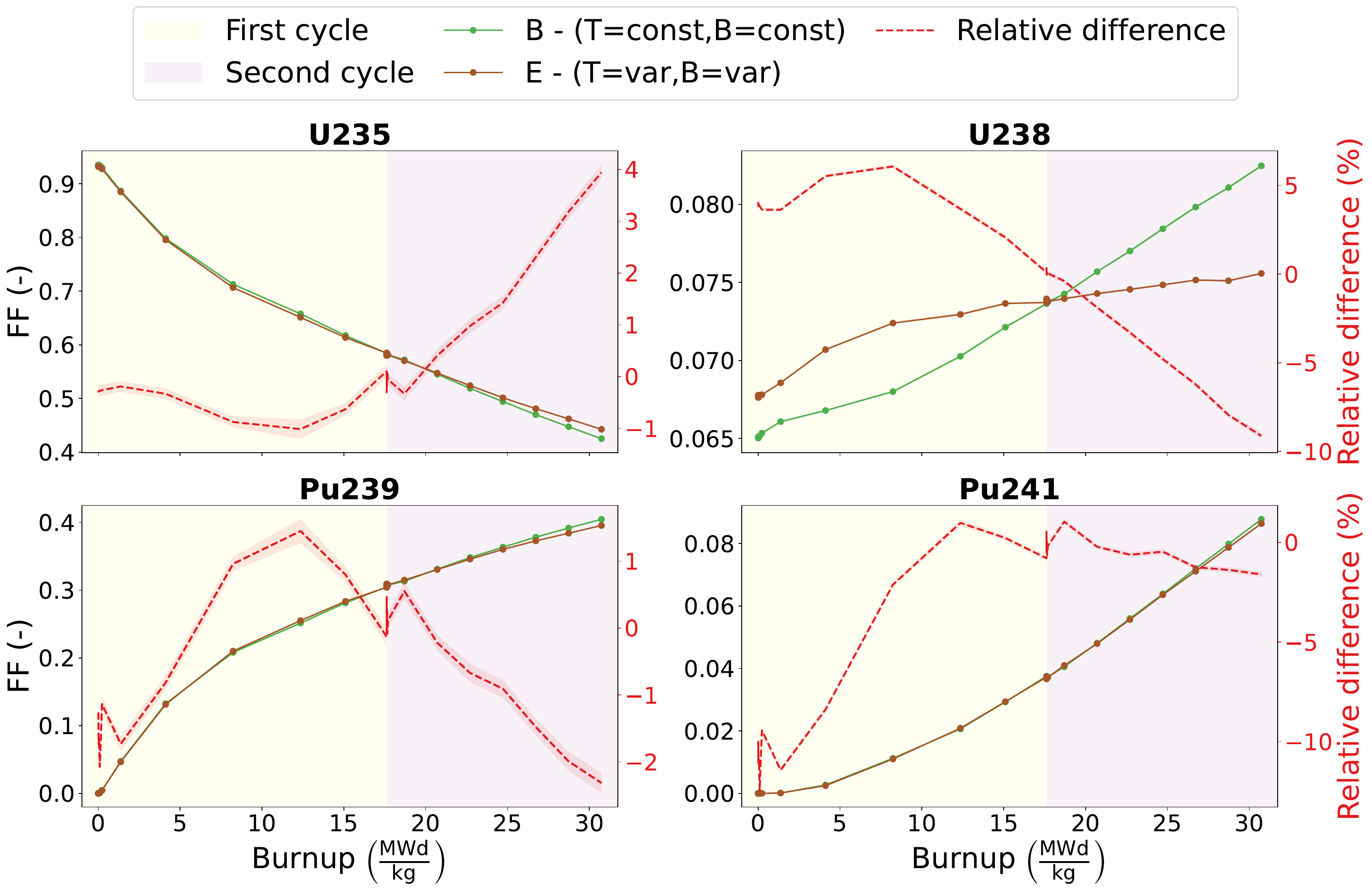}
    \caption{Comparison between fission fraction in simulation B (green line) and simulation E (brown line) and their relative difference (red dashed line). The combined effect of the boron treatment and spatial temperature shows a similar behaviour with respect to the B-D comparison, the combined effect matches with a superimposition effect within the design choices.}
    \label{fig:FF_95_BE}
\end{figure}

Finally, it is worth mentioning that the combined effect follows a superimposition effect: Figure~\ref{fig:FF_95_BE} compares the most simple case (B) with the most accurate one (E). Analyzing burnup data, the combined effect of temperature field and boron treatment is the sum of the single scenarios previously shown.

\end{document}